\documentstyle{mn}

\newif\ifAMStwofonts

\def\zsun{\thinspace\hbox{$\hbox{Z}_{\odot}$}}

\def\Zsun{\thinspace\hbox{$\hbox{Z}_{\odot}$}}

\ifoldfss
  \ifCUPmtlplainloaded \else
    \NewTextAlphabet{textbfit} {cmbxti10} {}
    \NewTextAlphabet{textbfss} {cmssbx10} {}
    \NewMathAlphabet{mathbfit} {cmbxti10} {} 
    \NewMathAlphabet{mathbfss} {cmssbx10} {} 
  \fi
  \ifAMStwofonts
    \ifCUPmtlplainloaded \else
      \NewSymbolFont{upmath} {eurm10}
      \NewSymbolFont{AMSa} {msam10}
      \NewMathSymbol{\upi}     {0}{upmath}{19}
      \NewMathSymbol{\umu}     {0}{upmath}{16}
      \NewMathSymbol{\upartial}{0}{upmath}{40}
      \NewMathSymbol{\leqslant}{3}{AMSa}{36}
      \NewMathSymbol{\geqslant}{3}{AMSa}{3E}

      \let\leq=\leqslant 
      \let\geq=\geqslant \let\ge=\geqslant
    \fi
  \fi
\fi 

\ifnfssone
  \newmathalphabet{\mathit}
  \addtoversion{normal}{\mathit}{cmr}{m}{it}
  \addtoversion{bold}{\mathit}{cmr}{bx}{it}
  \newmathalphabet{\mathbfit} 
  \addtoversion{normal}{\mathbfit}{cmr}{bx}{it}
  \addtoversion{bold}{\mathbfit}{cmr}{bx}{it}
  \newmathalphabet{\mathbfss} 
  \addtoversion{normal}{\mathbfss}{cmss}{bx}{n}
  \addtoversion{bold}{\mathbfss}{cmss}{bx}{n}
  \ifAMStwofonts
    \ifCUPmtlplainloaded \else
      \UseAMStwoboldmath
      \makeatletter
      \new@mathgroup\upmath@group
      \define@mathgroup\mv@normal\upmath@group{eur}{m}{n}
      \define@mathgroup\mv@bold\upmath@group{eur}{b}{n}
      \edef\UPM{\hexnumber\upmath@group}
      \new@mathgroup\amsa@group
      \define@mathgroup\mv@normal\amsa@group{msa}{m}{n}
      \define@mathgroup\mv@bold\amsa@group{msa}{m}{n}
      \edef\AMSa{\hexnumber\amsa@group}
      \makeatother
      \mathchardef\upi="0\UPM19
      \mathchardef\umu="0\UPM16
      \mathchardef\upartial="0\UPM40
      \mathchardef\leqslant="3\AMSa36
      \mathchardef\geqslant="3\AMSa3E

      \let\leq=\leqslant 
      \let\geq=\geqslant \let\ge=\geqslant
    \fi
  \fi
\fi 

\ifnfsstwo
  \DeclareMathAlphabet{\mathbfit}{OT1}{cmr}{bx}{it}
  \SetMathAlphabet\mathbfit{bold}{OT1}{cmr}{bx}{it}
  \DeclareMathAlphabet{\mathbfss}{OT1}{cmss}{bx}{n}
  \SetMathAlphabet\mathbfss{bold}{OT1}{cmss}{bx}{n}
  \ifAMStwofonts
    \ifCUPmtlplainloaded \else
      \DeclareSymbolFont{UPM}{U}{eur}{m}{n}
      \SetSymbolFont{UPM}{bold}{U}{eur}{b}{n}
      \DeclareSymbolFont{AMSa}{U}{msa}{m}{n}
      \DeclareMathSymbol{\upi}{0}{UPM}{"19}
      \DeclareMathSymbol{\umu}{0}{UPM}{"16}
      \DeclareMathSymbol{\upartial}{0}{UPM}{"40}
      \DeclareMathSymbol{\leqslant}{3}{AMSa}{"36}
      \DeclareMathSymbol{\geqslant}{3}{AMSa}{"3E}

      \let\leq=\leqslant 
      \let\geq=\geqslant \let\ge=\geqslant
    \fi
  \fi
\fi 

\ifCUPmtlplainloaded \else
  \ifAMStwofonts \else 
    \def\upi{\pi}
    \def\umu{\mu}
    \def\upartial{\partial}
  \fi
\fi

\begin{document}

\title{Bulges}
\author[M. Moll\'a, F. Ferrini and G. Gozzi]
     {Mercedes Moll\'a $^{1,3}, $ Federico Ferrini $^2$ and
      Giacomo Gozzi $^2$\\
     $^1$ Departement de Fisique, Universit\'{e} Laval, Chemin St-Foy,
     Quebec, Canad\'{a}\\
    $^2$ Dipartimento di Fisica, Universit\`a di Pisa, Piazza Torricelli 2, 56100 Pisa, Italy \\
    $^3$ Present address Departamento de F\'{\i}sica Te\'orica, C-XI,
  Universidad Aut\'{o}noma de Madrid, Cantoblanco 28049, Madrid, Spain}

\date{Accepted .... 1999.
      Received .... 1999 ;
      in original form .... 1999}


\pagerange{\pageref{firstpage}--\pageref{lastpage}}
\pubyear{1999}

\maketitle

\label{firstpage}

\begin{abstract}

We model the evolution of the galactic bulge and of the bulges of a selected
sample of external spiral galaxies, via the multiphase multizone evolution model.
We address a few questions concerning the role of the bulges within galactic
evolution schemes and the properties of bulge stellar populations.
We provide solutions to the problems of chemical abundances and spectral 
indices, the two main observational constraints to bulge structure.

\end{abstract}


\section{Introduction}

Bulges are the spheroidal component of spiral galaxies. Models for their
formation and evolution have been dominated in the past
by the strong parallelism with elliptical galaxies,
since several studies on
Galactic bulge stars (e.g. Whitford 1978; Frogel \& Whitford 1987;  Rich
1988: Frogel 1988) claimed that their photometric
properties were analogous to early-type galaxies and only
old and super-solar metal rich stellar populations seemed to exist in bulges.
Recently, younger stars were found in  Baade's window (BW, Holtzman et
al. 1993), suggesting a star formation history different from the one attributed to
elliptical galaxies, i.e. strongly concentrated in the early evolutionary stages
and turned off after the first Gyr.
Chemical abundances in the Galactic bulge have been revised 
(McWilliam \& Rich 1994; Minniti et al. 1995; Ibata \& Gilmore 1995a, 1995b;
Houdashelt 1996; Frogel, Tiede \& Kuchinski 1999, hereinafter FTK99)
and at present the iron relative abundance peaks at $\sim -0.3$ dex.
A similar result has been obtained by Bacells \& Peletier (1994) and
deJong (1996a,b) for external spirals, by estimating from new photometric data
a sub-solar mean metallicity $\rm <[Fe/H]> \sim -0.2,-0.3 $ dex, suggesting that
all bulges are very similar in colors and metallicities.
Kinematic observations show that bulges are
rotating much more rapidly than the bright elliptical galaxies,
suggesting that bulges are supported by rotation at variance to ellipticals.

The main scenarios of spiral galaxies formation constraint the bulge formation:
in the dissipative collapse model by Eggen, Lynden--Bell \& Sandage (ELS, 1962)
the evolution proceeds inwards and metallicity, age and kinematics are correlated.
Zinn (1985) proposed a merger scenario, with a halo formed by accretion of
small fragments, justifying the absence of a metallicity gradient in the halo.
The correlation of bulge and disc scale lengths and the similarity of inner disc and
outer bulge metallicities stimulated alternative models, such as a bar
producing a concentration of stars in the center
(Pfenninger \& Norman 1990; Pfenninger \& Friedli 1991) or a nuclear
starburst ejecting a giant gas bubble from which stars form
(Wada, Habe \& Sofue, 1995). Wyse
(1999) demonstrates that instabilities of purely stellar discs cannot
form bulges and that the metallicity distribution in MWG is not
consistent with a bulge built up from accretion of satellite galaxies
or clusters.

The ELS scenario seems to recover, supported also by recent observations
(Minniti, 1996a,b), and it  suggests that the bulge formed
through a dissipative collapse, resulting younger than the halo; FTK99  with
IR data show that the metallicity presents a strong radial gradient in the
inner bulge ($< 1.5$ kpc), in agreement with previous studies
(Terndrup, Frogel \& Whitford, 1990; 1991 and Minniti et al., 1995).
Zinn (1993) found that the inner old halo clusters ($\rm R < 6$ kpc)
show the signature of the dissipation, and that there is a smooth
transition from the halo to the disc globular clusters, suggesting
that they proceed from the same collapse.
Within this revival of the dissipative collapse,
Van den Bosch (1998, 1999) and  Elmegreen (1999) discussed
the bulge formation as due to  baryonic material from  halo or protogalaxy
settling in the center of the spheroid in a short time. The
natural trend is to form a bulge because the angular
momentum is low in the center, and it results older than the stellar disc.
In particular, Elmegreen evaluates the threshold density for the star formation
in turbulent conditions due to the gas accretion to form the
bulge: a starburst phase is established in a short time scale.
In both cases, the formation and evolution of bulges are different from
elliptical galaxies.

In the present paper we wish to address the following questions;
a few have already
challenged the modelers and others have not been previously discussed
in the literature:
\begin{itemize}
\item{} Is it needed to require the merging and/or accretion of
external material to reproduce the main observed characteristics of
bulges?

\item{} Is it possible to reproduce the observed properties of bulge
stellar populations?

\item{} How far goes the analogy with elliptical galaxies?

\item{} Which are the ages of bulges within the whole galactic evolution?

\end{itemize}

To afford these questions, we adopt the multiphase evolution model,
which assumes a dissipative collapse of
the gas from a protogalaxy or halo to form the bulge and the disc, in
ELS scenario. The model has been applied successfully to the Galaxy and
external galaxies (see papers by Ferrini and coworkers).

Moll\'{a} \& Ferrini 1995 (hereinafter MF) applied first the multiphase model
to study the evolution of the Galactic Bulge, considering it
as a natural extension of the disc, anyway with a separate evolution:
supernovae winds and no radial flows from the disc were assumed.
At variance with other models present in the literature, MF succeeded in
reproducing the metallicity data, revised at that date by McWilliam \& Rich
(1994). In the present approach, first we modify the modeling of the bulge,
by considering a core population in the central region. In this scenario, the nuclear
population is distinguishable from the bulge one as far as metallicity and dynamics
are concerned: nuclear starts are more metal--rich and have higher $\alpha$ elements
abundances. Furthermore, their motion must be supported by rotation, as predicted by
dissipative collapse models. Indications about a different kinematics of stars in the
galactic nucleus come from relatively large data sets of OH/IR stars (Lindquist et al. 1992)
and ordinary M giants (Blum et al. 1995) in the 5--200 pc region. These two samples appear to
trace different populations: while the M giants appear to joint smoothly with the kinematics
of other bulge tracers at larger radii, OH/IR stars have a lower velocity
dispersion ($\sigma_{OH/IR} = 50 - 100$ km/s against $\sigma_M > 100$ km/s) and
a very high rotational velocity ($V_{rot} > 100$ km/s at a distance of 100 pc from the
galactic center), against an almost negligible rotation of the M giants.
The OH/IR stars may be associated with the nuclear bulge star population.

We extend the discussion to a set of external galaxies, whose discs have been extensively
studied by our group (Moll\'a, Ferrini \& Diaz 1996, 1997).

The theoretical model is presented in \S 2; in \S 3 are the result of the
new bulge model. In \S 4 we analyze the chemical evolution results for all the
bulge sample, by including the study of possible trends with the galactic
properties (Hubble type, total mass). In the section \S 5, we
predict the values for the Lick spectral indices Mg$_2$ and Fe52 and
compare these predictions with observational data. Discussion and
conclusions are in \S 6.

\section{The Evolution Model}

In the multiphase approach the star formation is considered a two step process:
the diffuse gas forms molecular clouds, then stars form from cloud-cloud
collisions (spontaneous star formation) or by the interaction of
massive stars with molecular gas which induces  star
formation. We discussed extensively this approach in previous papers
(e.g. Ferrini et al. 1992) and fixed the relevant parameters
which define the coupling between the phases and the zones.
Interesting to our present paper is essentially the collapse time
scale $\tau_{coll}$, which defines the infall gas rate from the halo
to the secondary disc region or bulge. We already discussed  how
this varies with the galactic mass and with the galactic radius
(Ferrini et al. 1994).
Spontaneous star formation and cloud formation rates depend on the radial
distance, too, and on the Hubble type (Ferrini \& Galli, 1988).
The corresponding efficiencies $\epsilon_{\mu}$ and $\epsilon_{H}$
will be varied according to the prescription of Ferrini et al. (1994)
while induced star formation, due to
the interaction of massive stars with the diffuse gas
is considered as a local process, without any radial
dependence, and its efficiency has the same value used in our previous
disc models. Cloud formation and  star formation
by cloud-cloud collisions in bulge and core regions are enhanced in
agreement with our previous analysis of disc zones.

In the first application of the multiphase model to the bulge region
(MF), this was considered as the natural continuation of the disc,
as moving radially from the Sun to the Galactic Center. Hence 
the parameters for star and cloud formation were calculated following
their radial variation as in Ferrini et al. (1994). The first approach
gave results in good
agreement with observational data, showing that it does not appear necessary
to invoke a merger or an accretion from extragalactic matter to explain the
origin of the bulge.  However, observational data show that the bulge is
not the disc continuation toward
the Galactic center, but it represents a distinct region, as is it apparent
from IRAS and COBE results (Dwek et al. 1995).
In this paper we consider the galactic bulge as a distinct region respect to
the disc, formed by the infall of gas from the halo over the bulge,
without appreciable exchange of gas with the disc (Wyse \& Gilmore 1992).
We consider that the central galactic region is formed by three distinct
zones and corresponding populations: the bulge, the halo over the bulge
and the core.
The bulge is  modeled following Kent (1992) mass model as an oblate spheroid
flattened on the galactic plane with semi major axis $x=2$ kpc,
semi minor axis $z=1.5$ kpc, and with a mass $M_{B}=1.8 \times 10^{10}
M_{\odot}$. The core has the same shape of the bulge scaled to a semi
major axis of
500 pc. The three zones interact by means of gas infall toward the center
and through an outflow of gas driven by supernovae (SN) winds.
The time scales for collapse from halo to bulge ($\tau_{H}$) and from bulge
to core ($\tau_{B}$) have been chosen to reproduce the correct final masses
for the three zones. With this constraint,
we have used the values $\tau_{H}=0.7 $ Gyr and $\tau_{B}=10$ Gyr.
The collapse time we find in this way
for the halo is shorter but not very different from that found in MF by
extrapolating from the value in the solar neighborhood
($\tau_{H}\sim 1$ Gyr).
As in MF, a SN-driven wind occurs when the thermal energy produced by
combined SN explosions
is larger than binding energy, calculated as $E_{B}=GM(R)M_{gas}/R$, where
$M(R)$ is the total mass within radius $R$.

\begin{figure*}
\vspace*{12cm}
\includegraphics{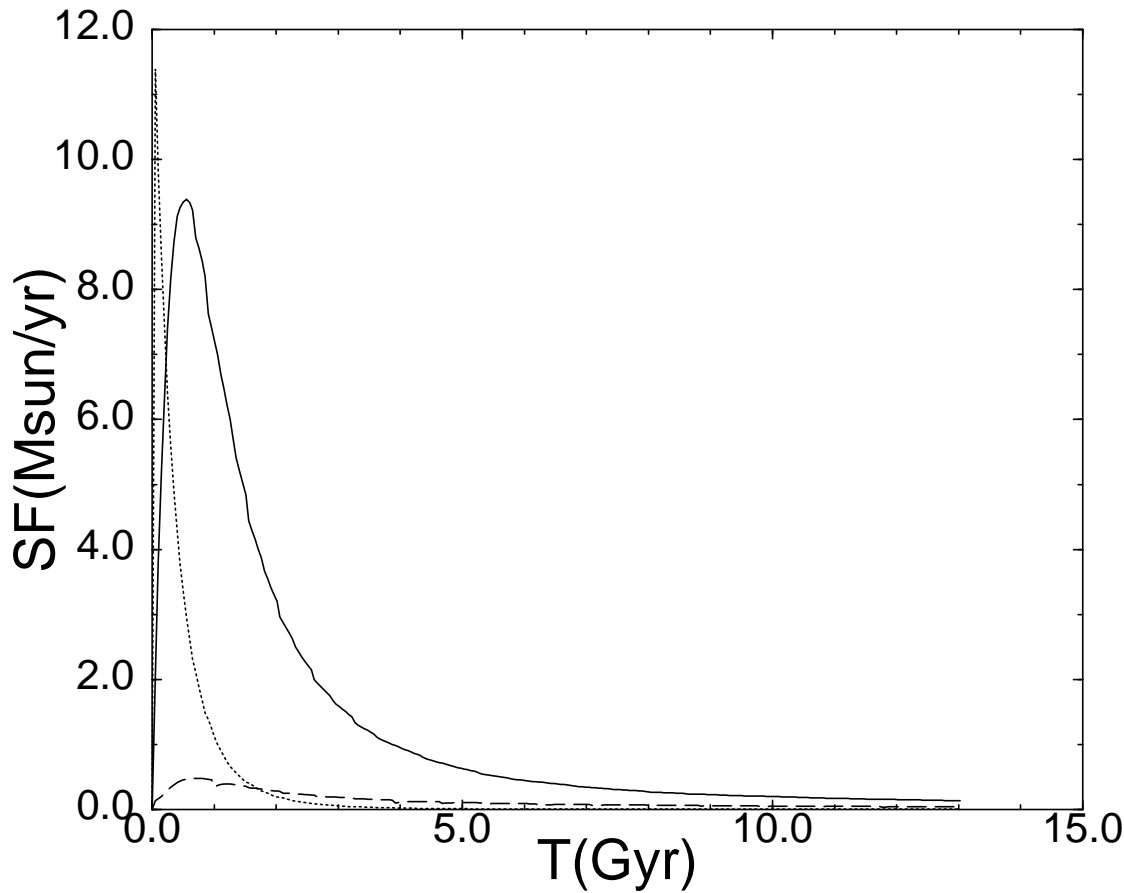}
\caption{SFR as a function of time for the three zones: bulge (solid line),
halo (dotted line) and core (long dashed line).
Line styles are the same for all figures.}
\end{figure*}

In comparison with MF we have updated $\alpha$ element nucleosynthesis
from massive stars, in the mass interval $8M_{\odot}\leq M
\leq100M_{\odot}$.  We use the computations of SN type II
nucleosynthesis by Woosley \& Weaver (1995) (WW95).  WW95 calculate
nucleosynthesis models for three different explosion energies and for
several initial abundances. Initial abundances do not affect very much
nucleosynthesis yields, while larger explosion energies increase them.
We adopt their model with intermediate explosion
energies (Model B) with initial solar abundances, to have a larger
number of stellar masses.

The model we are going to apply to the bulges of nine external galaxies
is the same used for the Galactic bulge, but simplified since we do not
include the core region. We assume a protogalaxy
with a sphere of primordial gas whose total mass $\rm M_{tot}$ is
calculated using the rotation curve $V(R)$:
$M(R)=2.32 \times 10^{5}\times V_{rot}(R)^{2}\times R$ (Lequeux, 1983).

\section{The Bulge of Milky Way Galaxy}
\subsection{Star Formation history}

Figure 1 shows the Star Formation Rate (SFR) in the three zones.  Star
formation in the halo over the bulge is similar to the star formation
in the halo over the solar region: it dominates at early times and is
essentially switched off after 2 Gyr.  The time scale for star
formation in the bulge is longer, approximately 5 Gyr, while the
maximum of star formation is reached at $t\simeq 0.6$ Gyr.  According
to our model there is a significant number of intermediate age ($<10$
Gyr) stars in the galactic bulge, in agreement with observational data
of Holtzmann et al. (1993).  In the core the maximum appears at the
same time, but the star formation rate reduces more slowly because of
the residual infall of gas from the bulge. 
 We can conclude that star formation starts in the external
region and continues mainly toward the center after about 0.5 Gyr; the
bulge formation occurred after the formation of the halo, in partial overlap
with the thick disc formation and almost completely before the 
thin disc produces efficiently stars.

\subsection{Element Abundances}

Figure 2a and 2b show respectively the time evolution of iron
abundance, [Fe/H] and of oxygen abundances, [O/H], for the three
zones.  The enrichment goes on during the collapse, then the relation
steepens as the radius decreases. The halo stars are very metal poor.
The average metallicity $\overline{[\rm Fe/H]}=-1.49$ is in good agreement
with observational data by Zinn (1985) for halo globular clusters
within a distance $R=3$ kpc ($\overline{[\rm Fe/H]}=-1.47$).

\begin{figure*}
\vspace*{8cm}
\includegraphics{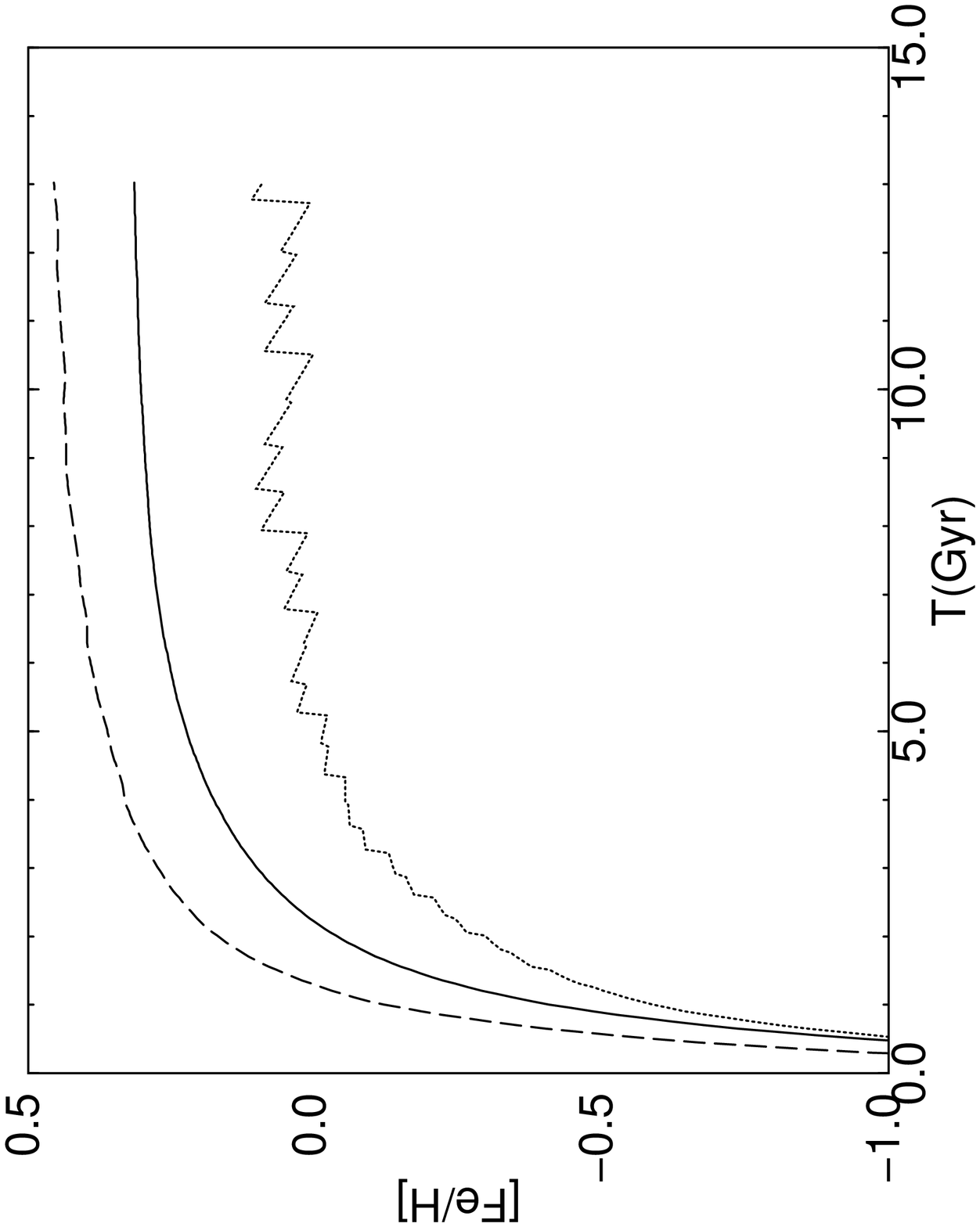}
\includegraphics{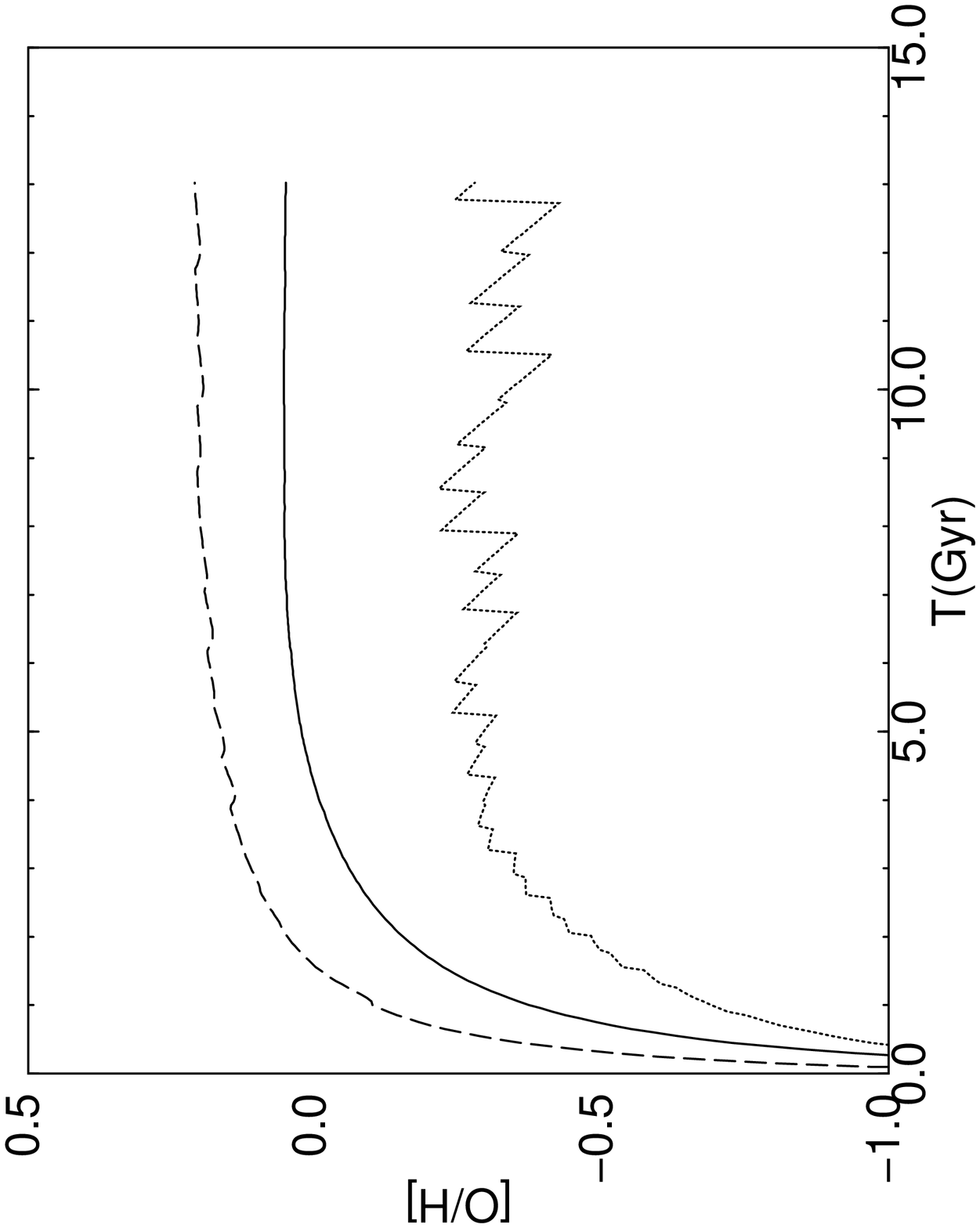}
\caption{Time evolution of ($a$) [Fe/H] and ($b$) [O/H] for the three
zones.}
\end{figure*}

The average metallicity $\overline{[\rm Fe/H]}$ for the bulge stars agrees
with the observational data by McWilliam \& Rich (1994 --hereinafter
McWR94) for a sample of K giants in the BW, as it will be discussed in
next section.

In Table 1 we present the main features of star formation in columns
2--4: SFR at 13 Gyr (SFR$_{now}$), the ratio of maximum to present SFR
($S_m/S_n$), the present mass fraction in each zone. In columns 5--9
are shown the main features of gas enrichment: the time at which solar
abundances are attained by iron and oxygen ($t_{Fe}$ and $t_O$
respectively), the present iron and oxygen abundances and the average
metallicity (iron abundance).

\begin{table*}
\begin{minipage}{50mm}
\caption{SFR and metal enrichment}
\begin{tabular}{@{}lccrrrrrr@{}} 
 & & & & & & & & \\
 & {SFR$_{n}$}  & {$S_{m}/S_{n}$} & {Mass} &
 {$t_{Fe}$} & {$t_{O}$}  & {[Fe/H]$_{n}$}&   {[O/H]$_{n}$}&
{$\overline{[\rm Fe/H]}$} \\ 
 & $M_{\odot}/yr$ & & \% & Gyr & Gyr & & & \\  
 & & & & & & & & \\
halo & 7$\times 10^{-4} $ & 1.6 $\times 10^4$ & 10 & 4.5  &
$-$ & 0.00 &
$-0.30$ &  $-1.49$ \\
bulge & $0.13$ &70 &83  & 2.0 & 4.1 & +0.30 & $+0.05$ &
$-0.26$ \\
core & 0.04  &15 &7  & 1.2 & 1.5 & $+0.45$ & $+0.20$ & $+0.22$ \\
 & & & & & & & & \\
\end{tabular}
\end{minipage}
\end{table*}

The abundance of $\alpha$ elements is of particular utility to
understand the formation time scale of the galactic bulge.  In Figures
3a-3d we show our model results for [O/Fe], [Mg/Fe], [Ca/Fe] and
[Si/Fe] versus [Fe/H] for the three zones compared with the data by
McWR94 for K giants in the BW. From the Figures 3 we can see that core
stars present enhanced abundances of $\alpha$ element compared with
bulge ones. Observational data for O, Ca and Si are very well
reproduced. On the contrary, our model is not able to reproduce
observational abundance for Mg. The abundance ratio [Mg/Fe] found by
McWR94 is enhanced by $\sim0.3$ relative to solar over almost the
entire [Fe/H] range. Only the two stars with the smallest [Fe/H]
ratios are reproduced by our model. This underestimation is probably
due to the SN II nucleosynthesis model we have used. Thomas, Greggio
\& Bender (1998) compare SN II nucleosynthesis models by WW95 with
that of Thielemann, Nomoto \& Hashimoto (1997), focusing on the
[Mg/Fe] ratios; they find that the last calculation model leads to an
enhancement of Mg abundance by 20 per cent relative to WW95 models.
To reproduce the Mg abundance, the SN II nucleosynthesis would be two
times higher than predicted by WW95 model. We can conclude that no
model available today can reproduce Mg abundance observed in the bulge
K stars.

\begin{figure*}
\vspace*{16cm}
\includegraphics{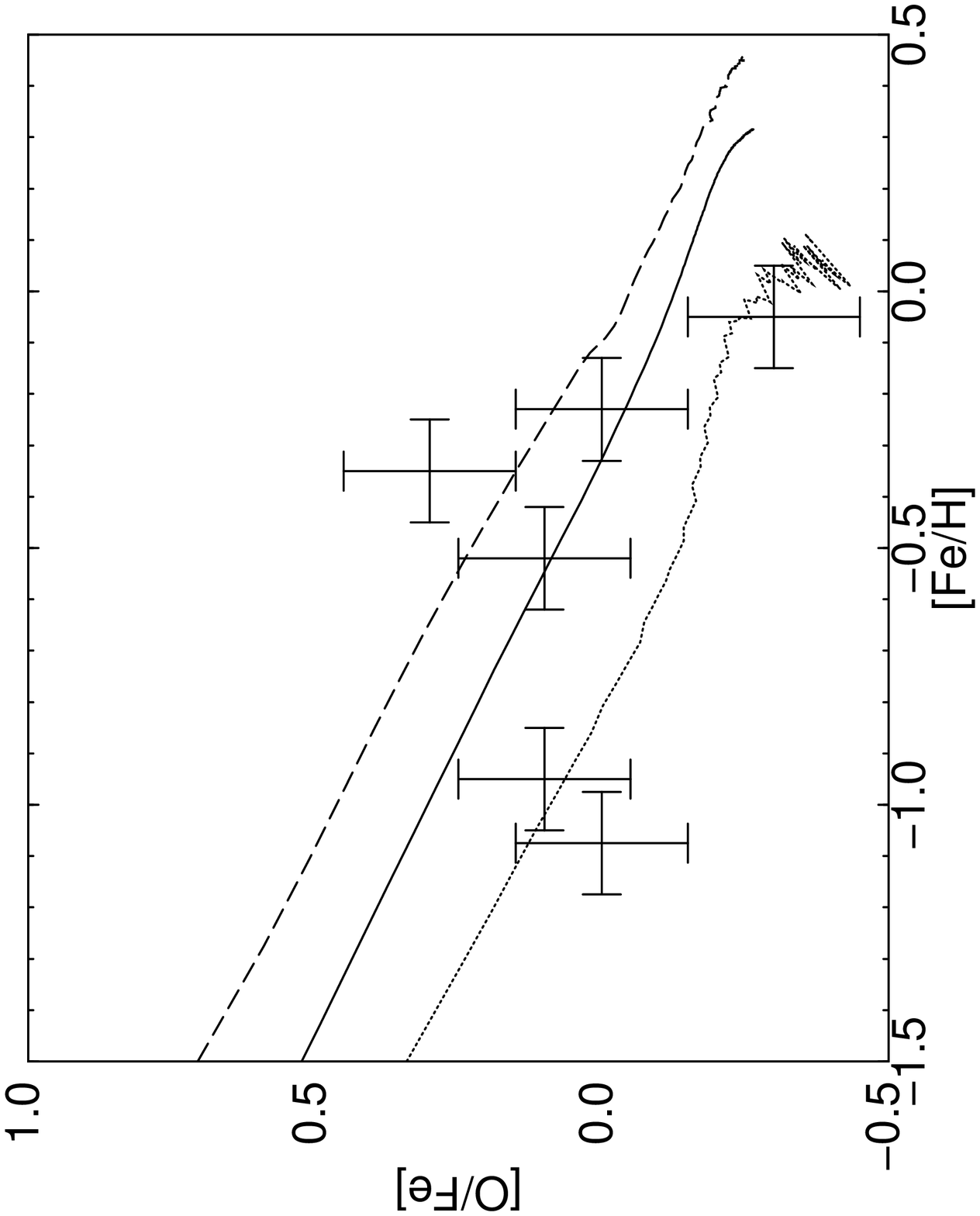}
\includegraphics{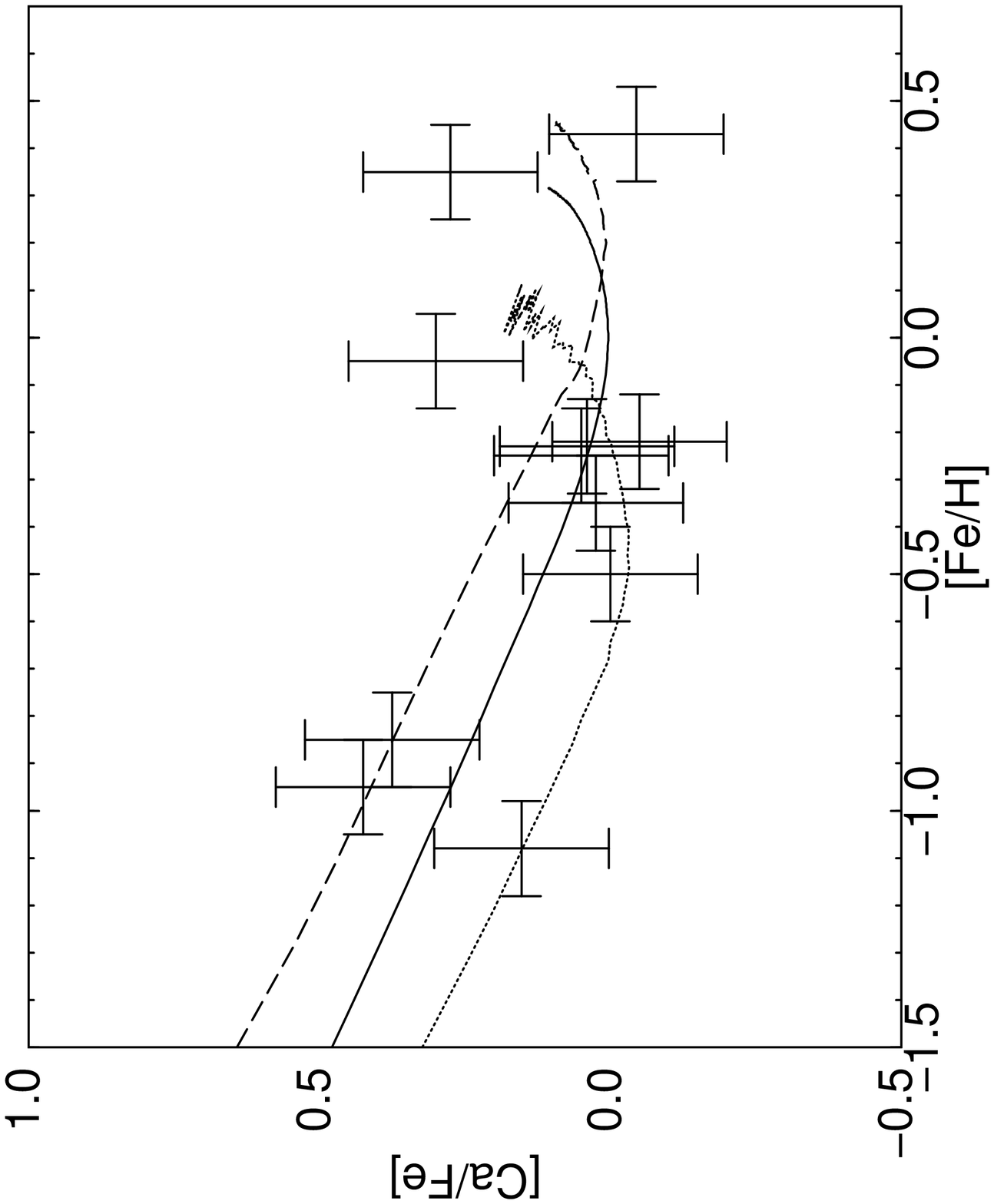}
\includegraphics{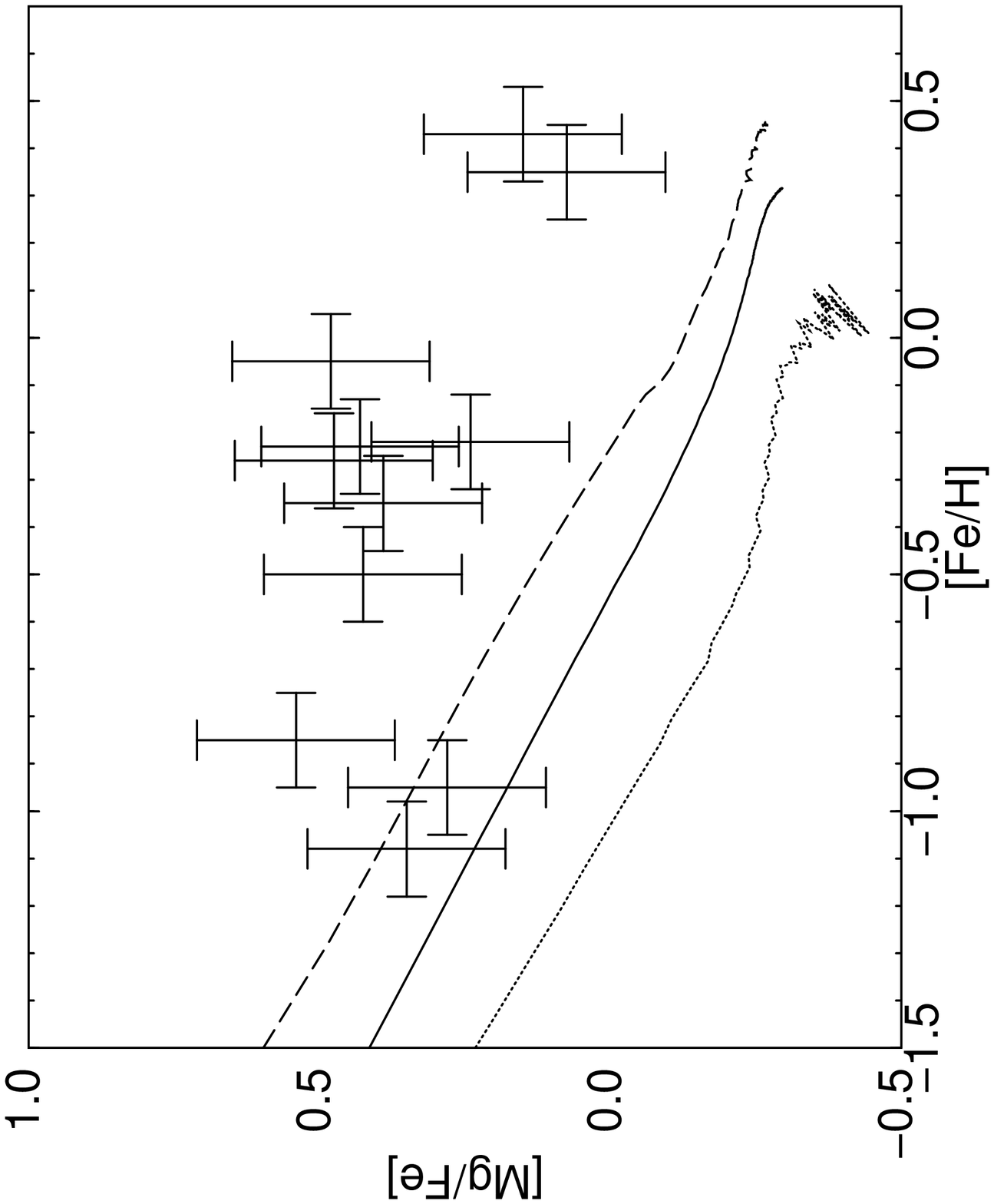}
\includegraphics{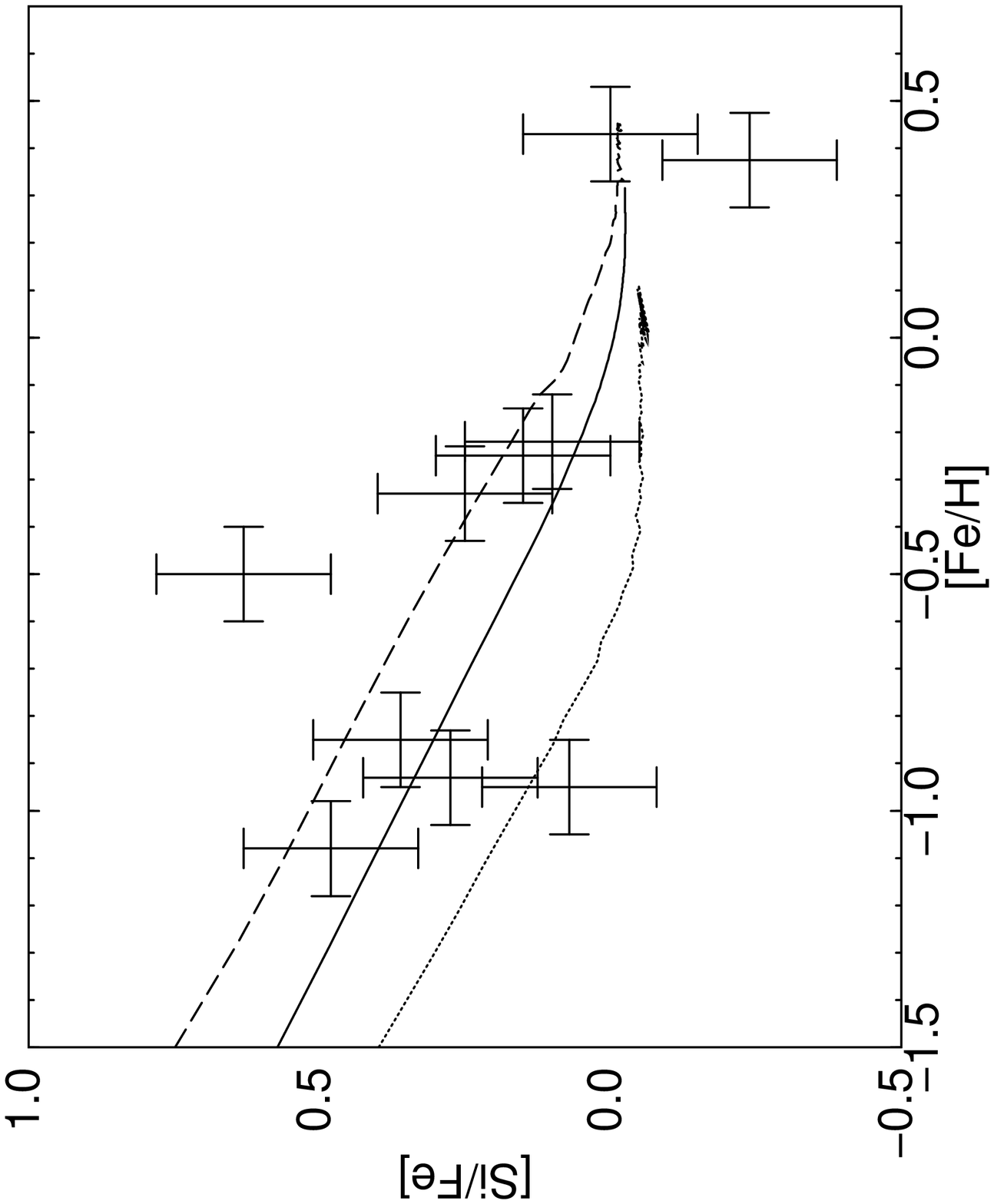}
\caption{[$\alpha$/Fe] versus [Fe/H] for the three zones.
($a$) oxygen, ($b$) magnesium, ($c$) calcium ($d$) silicon.
Data are from Mc William \& Rich (1994).}
\end{figure*}

Figure 4a shows the model result for Mg abundances compared with the
data by Sadler, Rich \& Terndrup (1996) (SRT96). The displayed data
represent the average values of [Mg/Fe] computed dividing the stars
sample into four bins.  Model results are lower but still consistent
with SRT96 observational data.  In Figure 4b we shows our model
results for [(C+N)/Fe] ratios versus [Fe/H] compared with SRT96 data,
collected as in Figure 4a.  The observational data seem well
reproduced, even if the large uncertainty for abundance ratio
($\Delta_{[C+N/Fe]}=0.4$ dex) reduces the validity of this result.

\begin{figure*}
\vspace*{8cm}
\includegraphics{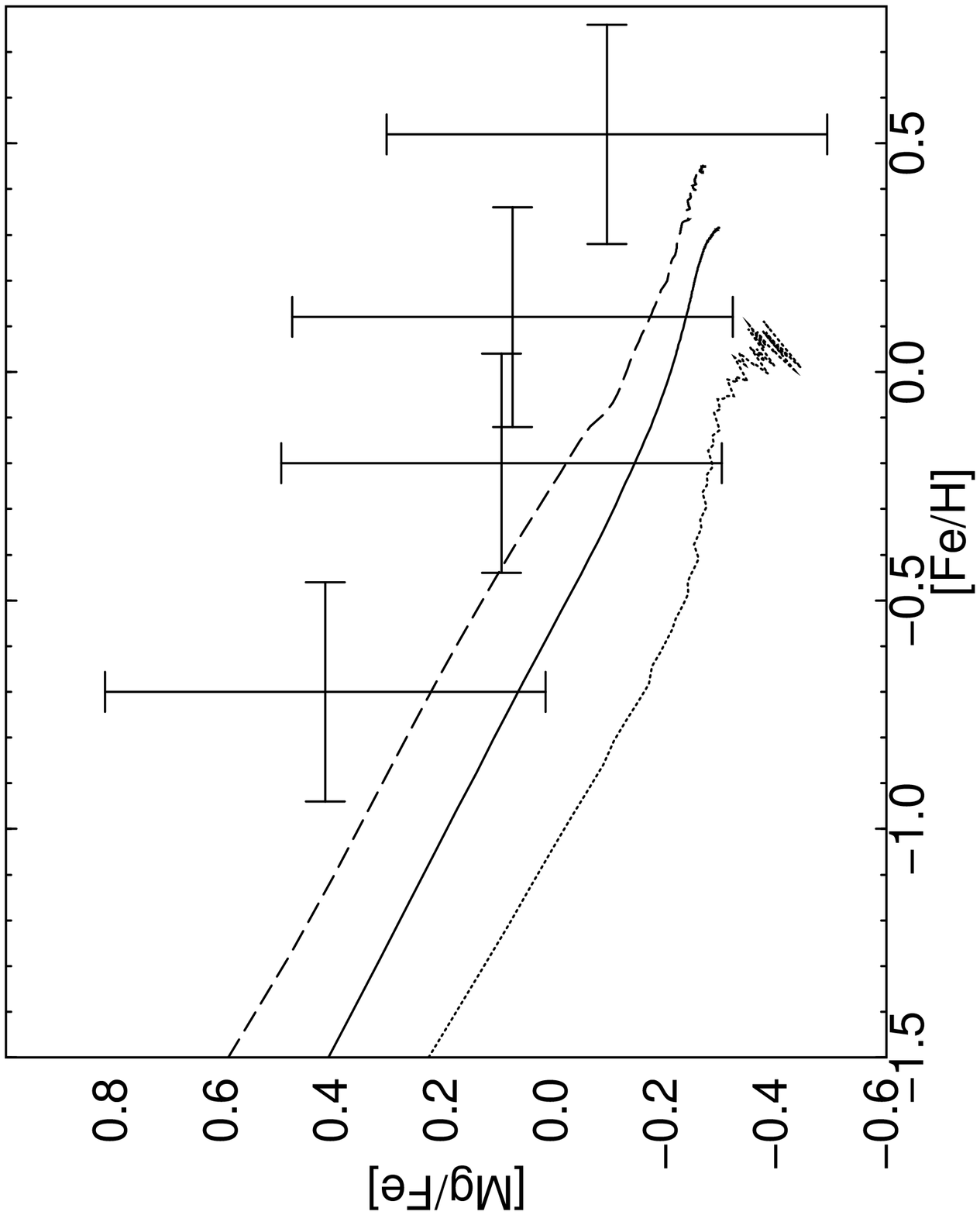}
\includegraphics{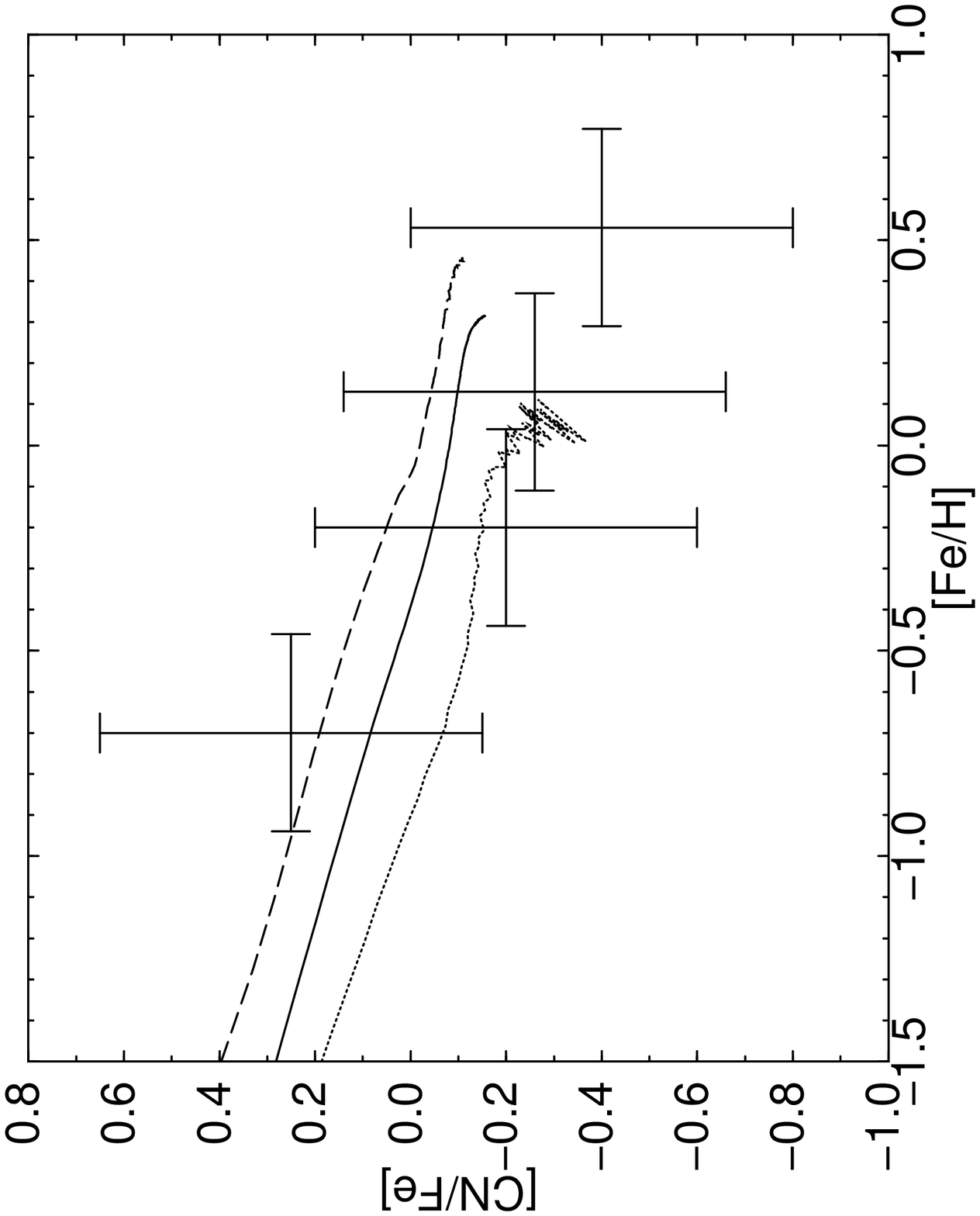}
\caption{[Mg/Fe] ($a$) and ($b$) [C+N/Fe] versus [Fe/H] for the three zones.
 Data are  from SRT96.}
\end{figure*}

\subsection{Metallicity distributions}

The metallicity distribution of K giants is an important observational
constraint to understand the sequential time formation of the galactic
bulge. Our model gives the time evolution of the total Star Formation
Rate $\rm \psi(t)$. To obtain the theoretical number of stars in the K
giants phase from $\rm \psi(t)$ we used the isochrones from the FRANEC
code  and the Initial Mass Function adopted in this series of
papers, derived by Ferrini, Palla \& Penco (1990). All the isochrones
correspond to an helium abundance $Y=0.27$, according to Minniti
(1996a), who finds $Y=0.28\pm0.02$ with a photometric determination in
the IR band of bulge stars. We used two different total metallicity:
Z=0.01 for stars with age $t<2$ Gyr and Z=0.02 for $t>2$ Gyr. In
Figure 5a-5c we show the theoretical K giants metallicity distribution for the three
zones. All the distributions are normalized to unity. Moving from the
halo to the core an increase of $\overline{[\rm Fe/H]}$ and a decrease of
standard deviation $\sigma_{[Fe/H]}$ are evident. In Figure 6a we
present the frequency histogram of Rich (1988) solution 1 [Fe/H]
corrected by the regression relation of McWilliam \& Rich (1994) and
the model result for the bulge region. Observations and theory agree
reasonably in the range $[Fe/H]>-1.25$. For lower iron abundances the
model predicts too many stars, as it happens with the G dwarf problem
in the solar region. Pardi, Ferrini \& Matteucci (1995) 
take into account the effect of the thick disc in the Galactic evolution;
the presence of the this intermediate phase of accretion to the thin disc
allows the solution of the G dwarf problem. We may easily imagine that
the introduction of a similar intermediate zone modifies the
theoretical distribution on the low metallicity end.

By comparing McWR94 observational data with theoretical bulge K giants
distribution of metallicity, we consider implicitly that every star in
BW belongs to the bulge population, without contaminations from halo
and core populations. There are anyway indications of the existence of
a contribution of halo star population in the BW: Rich (1990) finds a
correlation between kinematics and metallicity in a sample of K giants
in the BW: giants with [Fe/H] $< -0.3$ have higher velocity dispersion
than giants with [Fe/H]$>-0.3$. The velocity dispersion of the metal
poor K giants is consistent with their belonging to the same
$r^{-3.5}$ spheroid as globular cluster and RR Lyrae. Furthermore, in
this paper we consider the existence of a third stellar population,
the core one, distinct from bulge population in its metallicity
distribution. With this assumption a fraction of giants in the BW,
inside 550 pc from the galactic center, belongs to core population.
So we can conclude that in the BW there are not only present stars
belonging to the bulge population, but there must exist an
overlap of the three populations.

\begin{figure}
\vspace*{19cm}
\includegraphics{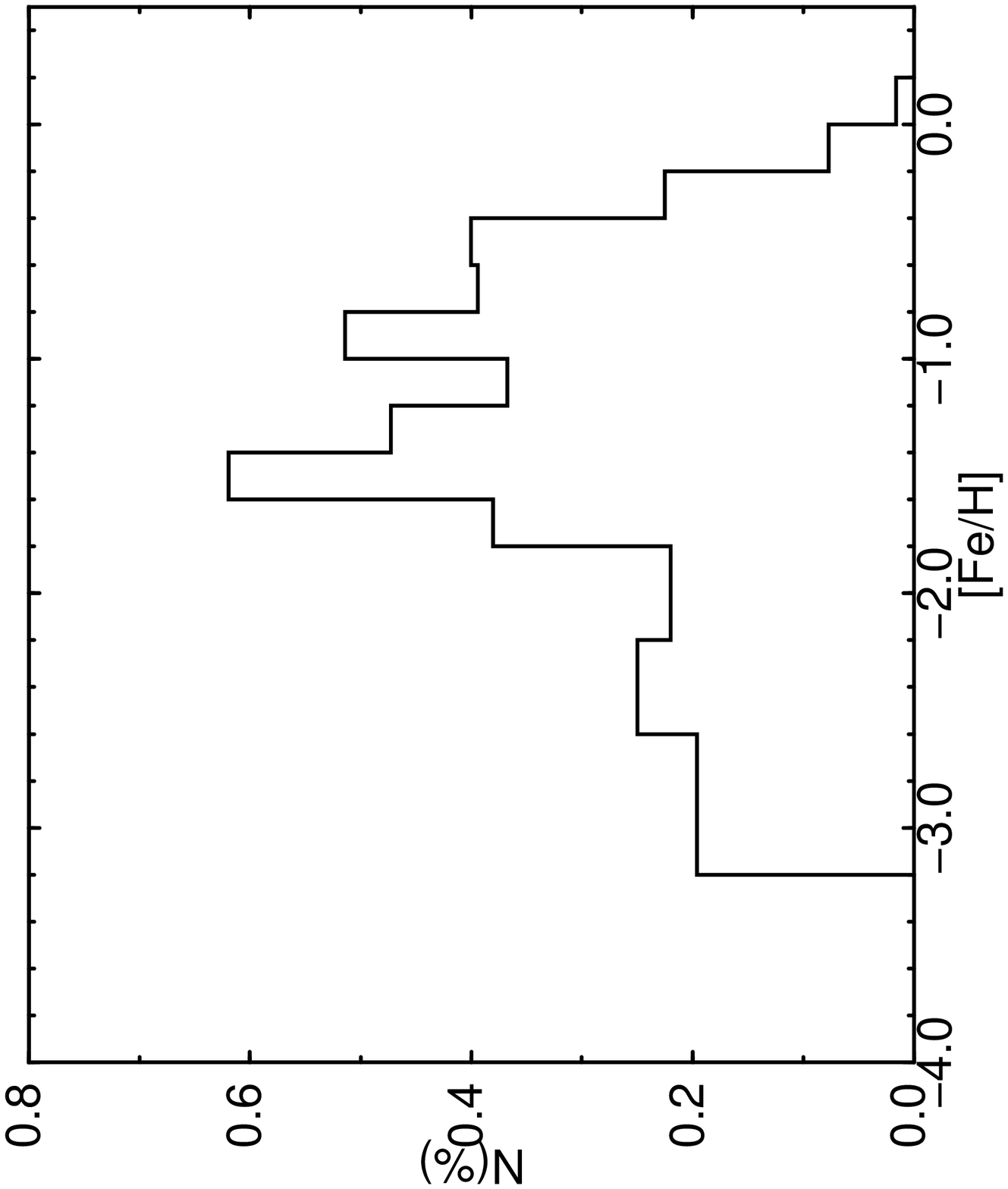}
\includegraphics{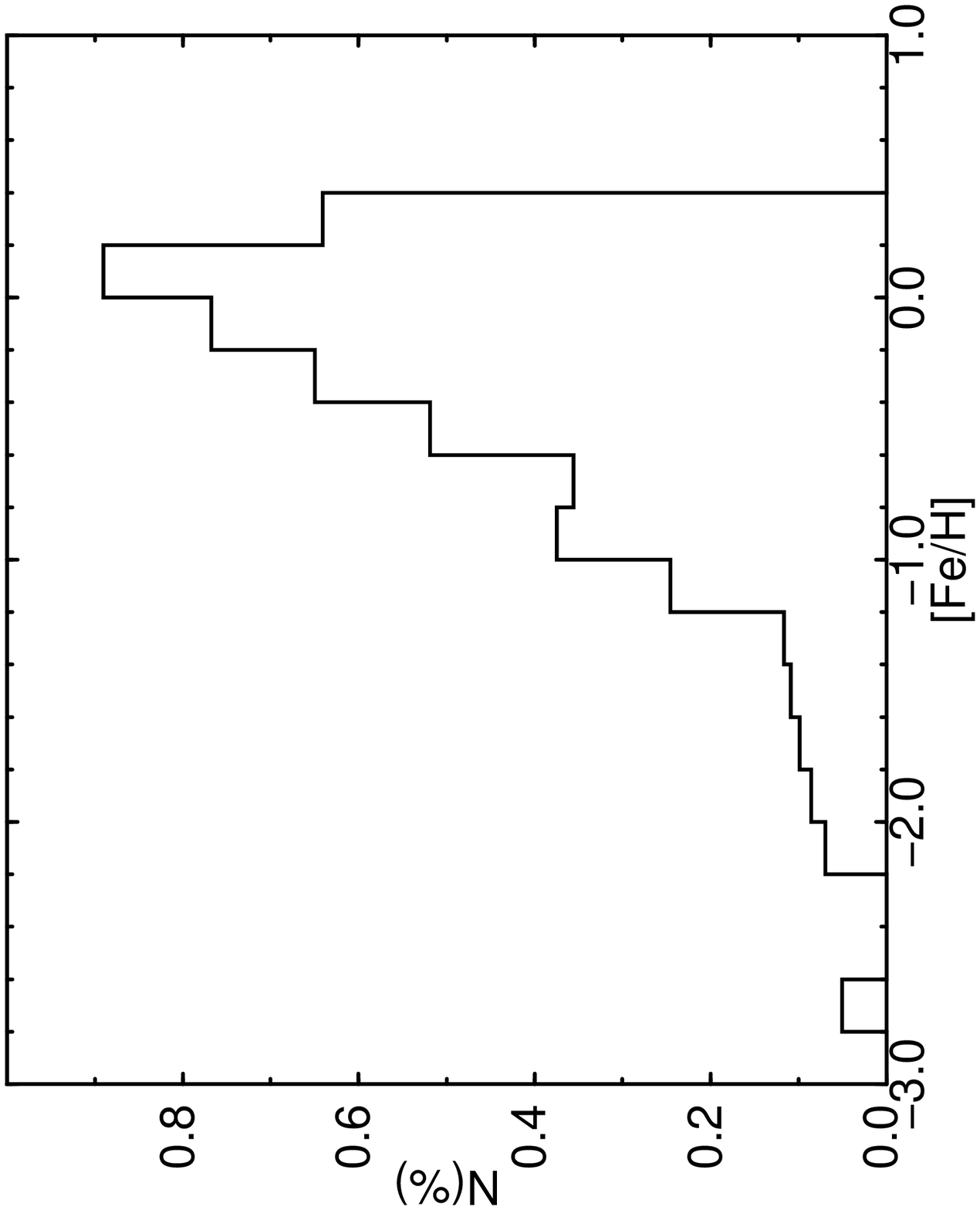}
\includegraphics{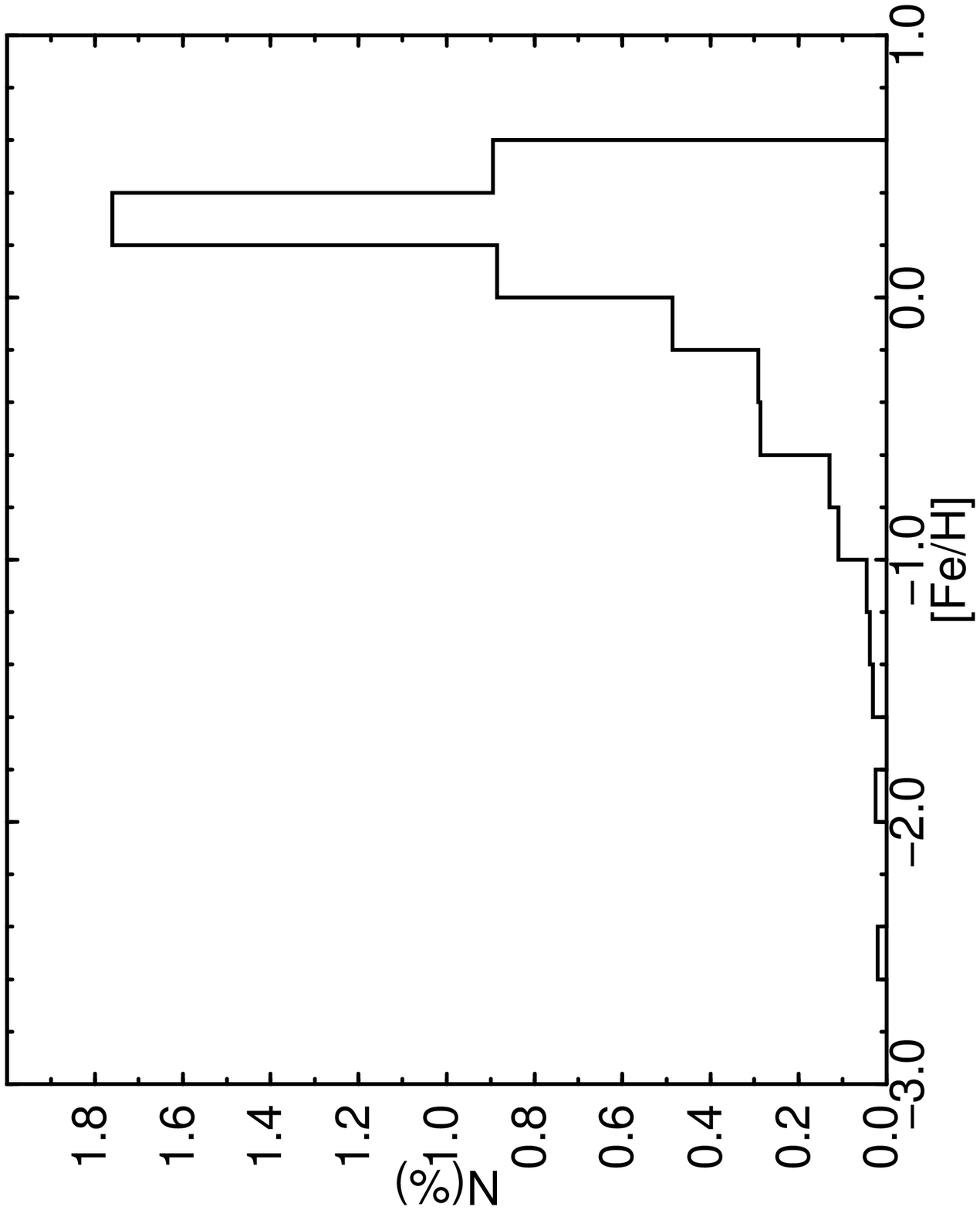}
\caption{K giants metallicity distribution. ($a$) halo, ($b$) bulge, ($c$)
core.}
\end{figure}

\begin{figure}
\vspace*{13cm}
\includegraphics{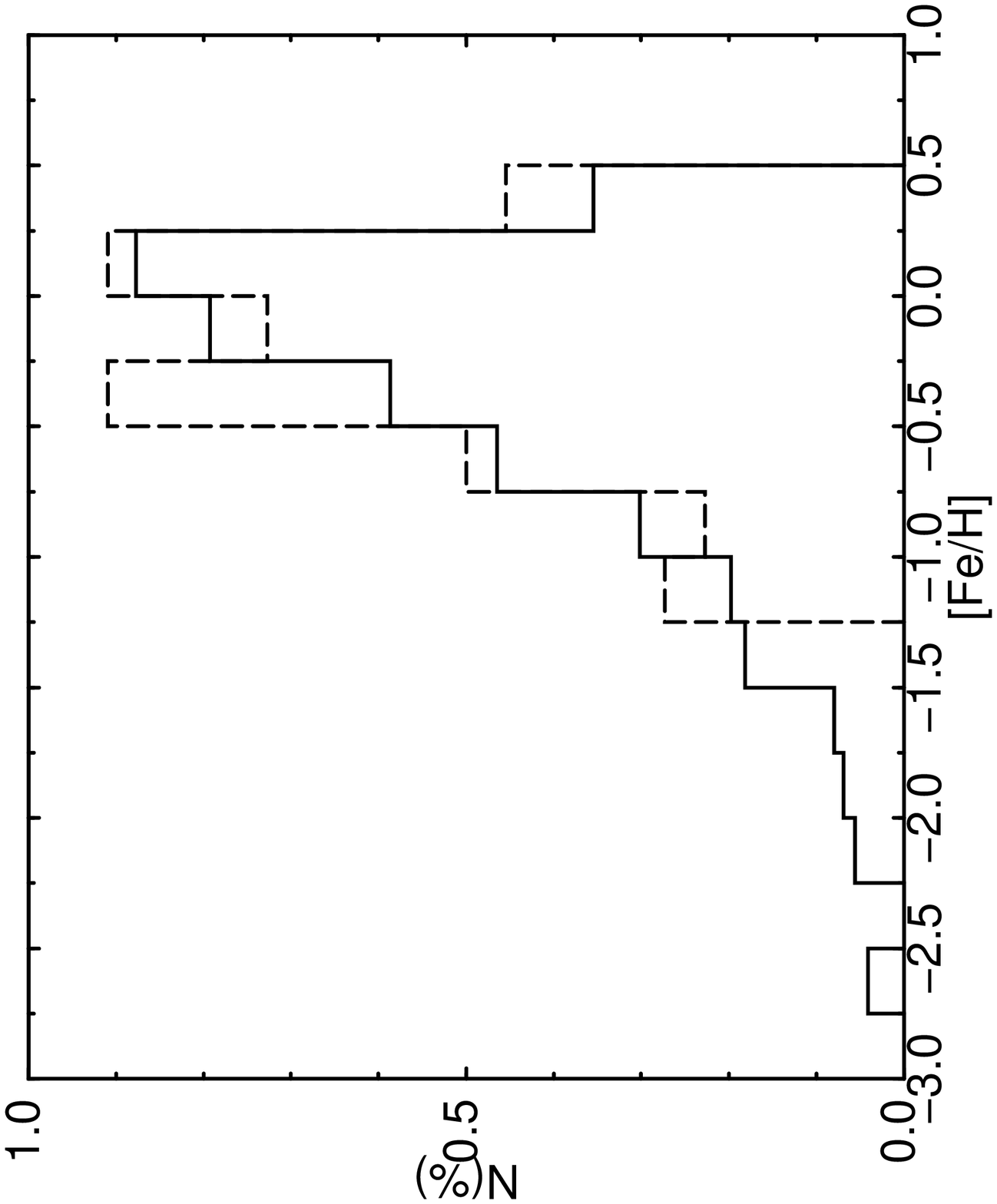}
\includegraphics{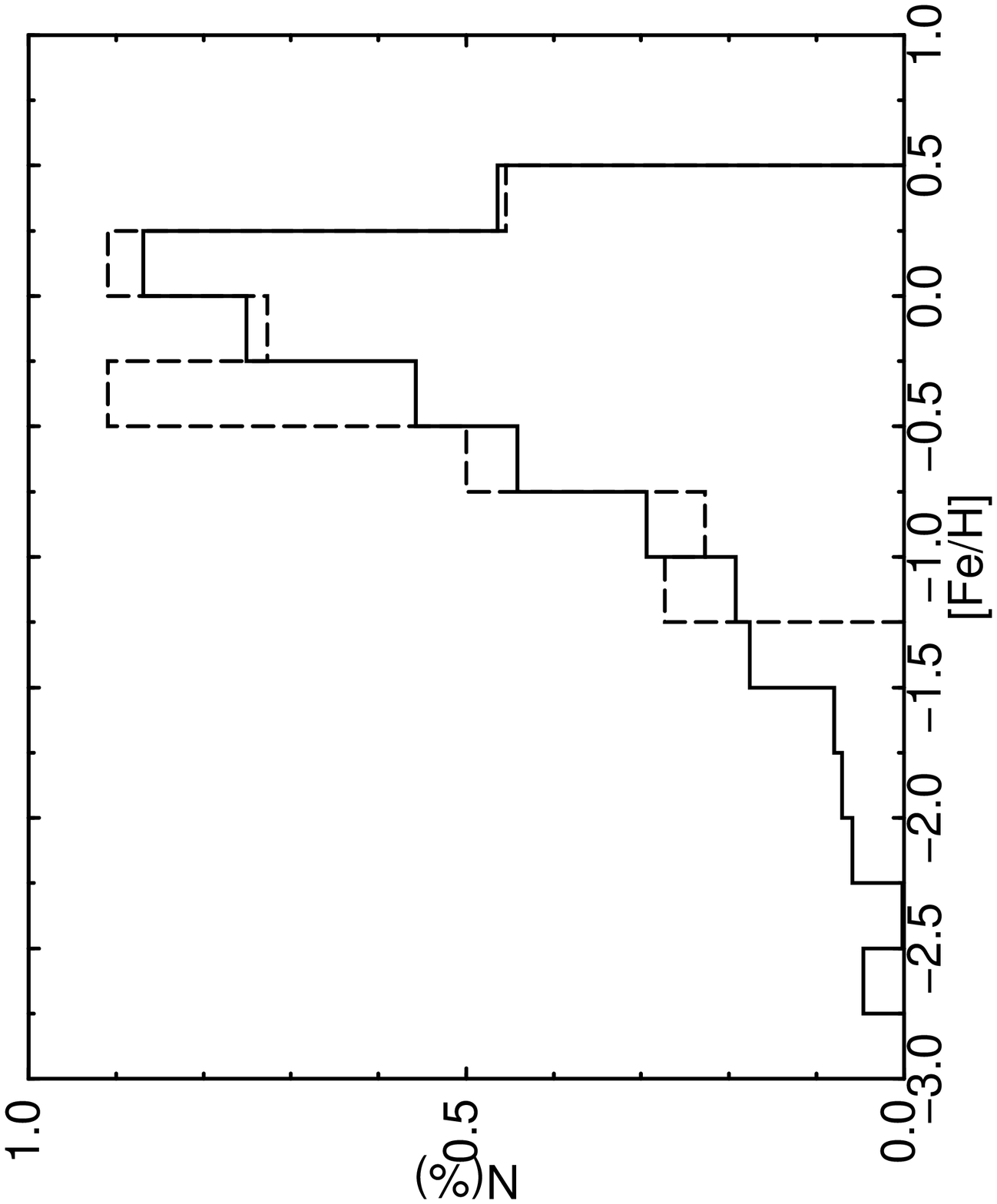}
\caption{K giants metallicity distribution by Mc William \& Rich (1994)
(dashed) compared with ($a$) bulge and ($b$) combination 1 theoretical
ones (solid).}
\end{figure}

\begin{table}
\caption{K giants distributions: mean metallicity and standard deviations}
\begin{tabular}{@{}llr@{}} 
 & &  \\
 & {$\overline{[\rm Fe/H]}$} & {$\sigma_{[\rm Fe/H]}$} \\ 
 & &  \\
halo  & $-$ 1.60 & 0.85 \\
bulge & $-$ 0.52 & 0.65 \\
core  & $-$ 0.07 & 0.45 \\
comb. 1 & $-$ 0.47 & 0.60 \\
comb. 2  & $-$ 0.42 & 0.60 \\
comb. 3  & $-$ 0.91 & 0.80 \\
comb. 1 truncated & $-$ 0.20 & 0.40 \\
comb. 2 truncated & $-$ 0.18  & 0.45 \\
McWR94 (sol. 1) & $-$ 0.25 & 0.40 \\
SRT96 & $-$ 0.11 & 0.45 \\
M95 (F588) & $-$ 0.60 & 0.70 \\
M95 (M22)  & $-$ 0.60 & 0.70 \\
 & &  \\
\end{tabular}
\end{table}

In order to compare correctly theoretical results with observational
data, we have to estimate the contributions from the different
galactic components. To compute the contribution of the three stellar
populations by varying the distance from Galactic Center we can use the
corresponding mass distribution models. We have considered the mass
model G2 of Dwek et al. (1995) for bulge star distribution; the halo
star contribution has been modeled following an axisymmetric power law
following Minniti et al. (1995). There are not available accurate core stars
mass models; because of the stronger concentration, the contribution
of core star population to the total star number can be relevant only
in the BW region, up to 550 pc from the galactic center. So we have
considered two combinations for the BW, differing in the core stars
contribution.  In combination 1, bulge stars represent 85\%, halo
stars 5\% and core stars 10 \% of the total stars.  In combination 2
we have considered a larger contribution of core stars, choosing
respectively 75\%, 5\% and 20\%, for bulge, halo and core.
Figure 6b shows McWR94 metallicity histogram compared with combination
1.  Combination 1 better reproduces observational data, even if for
iron abundances lower than [Fe/H]=-1.25 our model still predicts too
many stars.

\begin{figure}
\vspace*{13cm}
\includegraphics{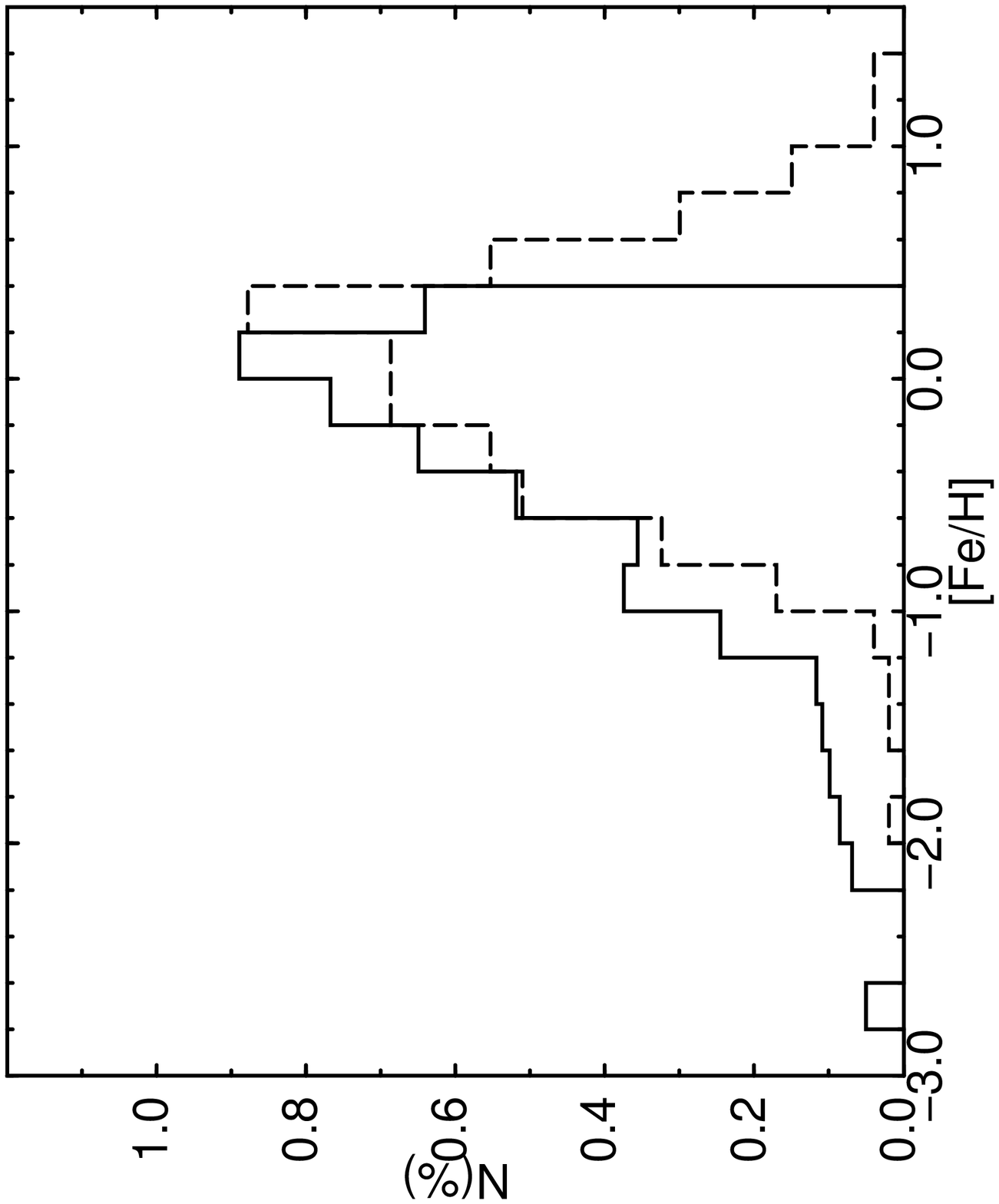}
\includegraphics{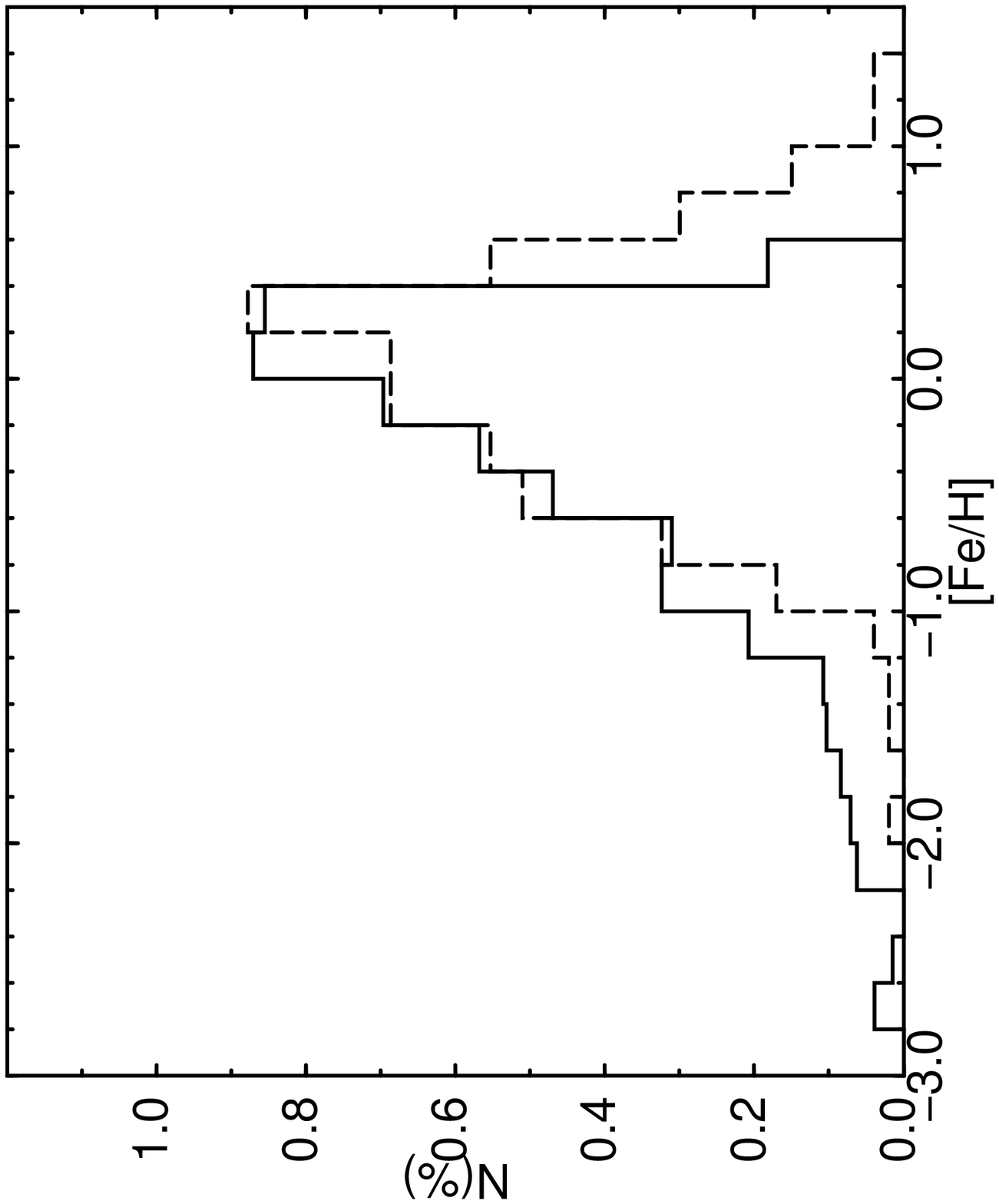}
\caption{K giants metallicity distribution by SRT96 (dashed line) compared
with ($a$) bulge and ($b$) combination 2 theoretical ones.}
\end{figure}

In Figures 7a and 7b, we show the metallicity distribution of pure
bulge and combination 2 respectively compared with SRT96 data.  SRT96
data are better reproduced by combination 2, in which core stars
contribution is 20\%.  From Table 2 it is clear that the choice of
combinations satisfies better the fitting of distributions. In Table 2
we insert also the properties of the combinations truncated to the low
metallicity end, to simulate the presence of the thick disc.

\begin{figure}
\vspace*{13cm}
\includegraphics{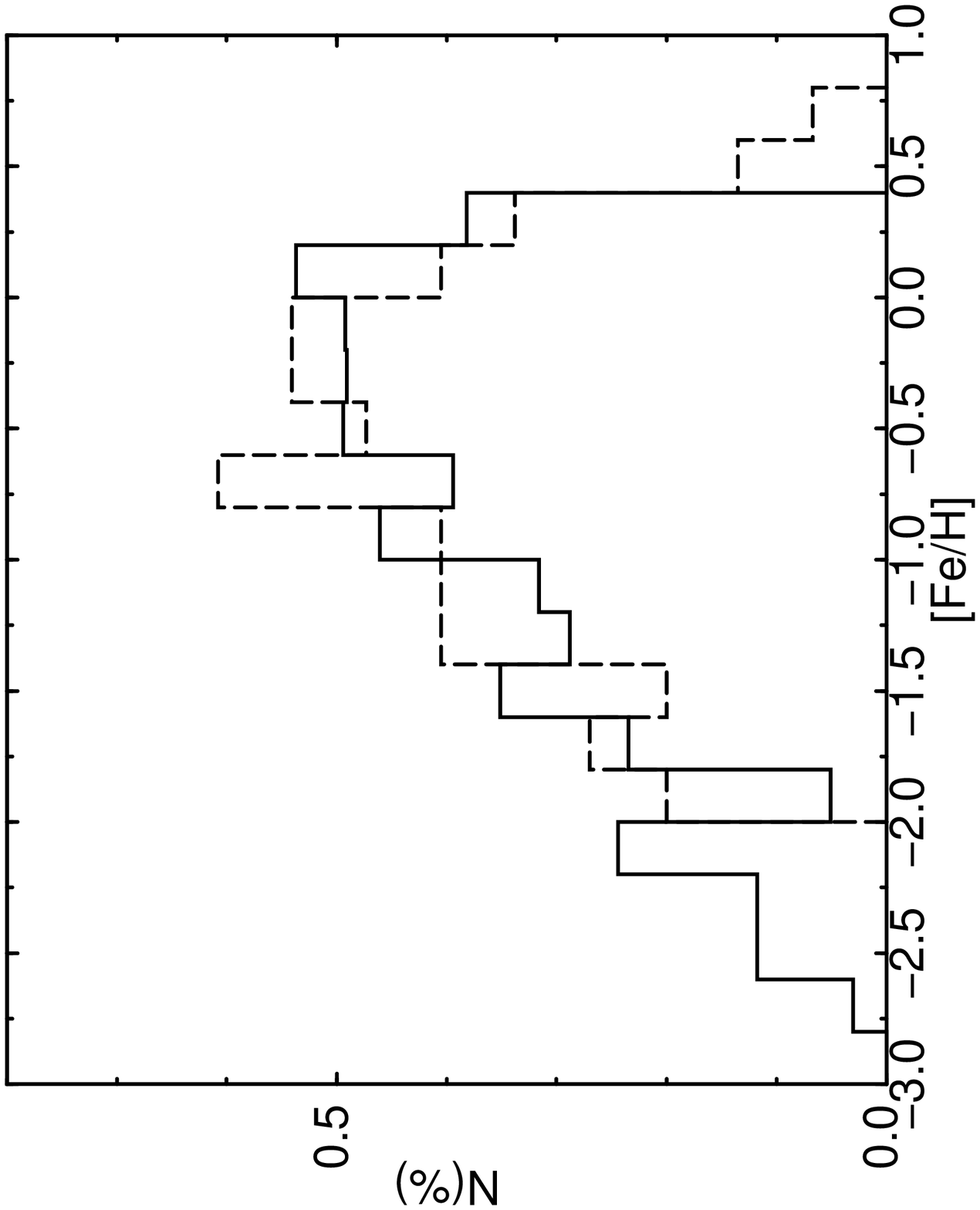}
\includegraphics{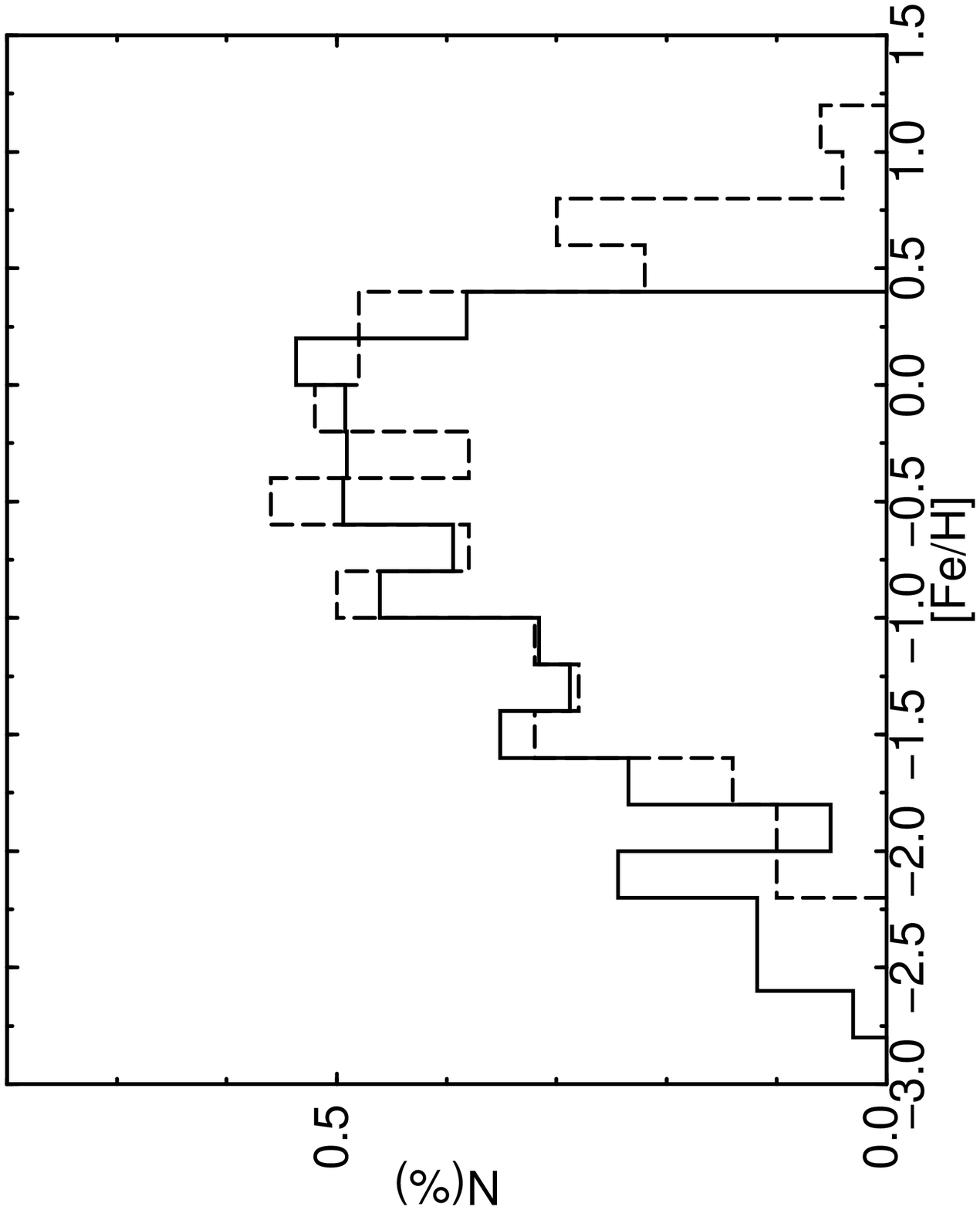}
\caption{K giants metallicity distributions by Minniti (1995) (dashed line)
compared with combination 3. ($a$) globular cluster M22, ($b$) field F588.}
\end{figure}

Figures 8a and 8b show Minniti et al. (1995) metallicity distribution of star
samples in F588 bulge field at $l,b=(8^{\circ},7^{\circ})$ (8a) and in
globular cluster M22 at $l,b=(8^{\circ},7^{\circ})$ (8b),
corresponding to distances $R=1.5-1.7$ kpc from Galactic Center.
Observational distributions are compared with theoretical data of
combination 3 (halo stars 30\%, bulge stars 70\%)
which corresponds in our simple scheme to these bulge
regions. The observational data are well reproduced by combination 3
in the metallicity range $ -2.0\leq$[Fe/H] $\leq +0.4$. It is still
present an overestimation for star with [Fe/H] $\leq -2$. On the other
hand the absence of theoretical stars with [Fe/H]$\geq +0.4$ can be
explained considering that the scatter in the metallicity
observational determinations increases for the most metal rich giants.
The fact that the errors increase with metallicity to
$\Delta_{[Fe/H]}=0.4$ for the more metal rich stars cast doubts on the
reality of the values for the most metal rich stars. Indeed, it is not
clear whether there are any stars in the samples with [Fe/H]$\geq
+0.5$ even in the more central zone or core (Jablonka
et al. 1996; Davidge, 1998).

\subsection{Metallicity gradient}

The determination of an abundance gradient in the central
regions of the Milky Way is important for testing models of Galaxy
formation. A metallicity gradient is predicted by dissipative
collapse models. We have calculated the theoretical
metallicity gradient comparing the average $\overline{[\rm Fe/H]}$ of K
giants distributions at different distances from the Galactic center.  We
use combination 1 for giants in the BW ($R=550$ pc) and combination 3
to reproduce giants distribution at $R=1.5$ kpc. We find the existence
of a metallicity gradient d[Fe/H]/dr $=-0.4$ dex/kpc in the bulge
region ($0.5 \leq R \leq 1.5$ kpc). We have calculated the metallicity
gradient in the core region with the assumption that near the Galactic
center, the core star population dominates. We find the existence of a
metallicity gradient d[Fe/H]/dr$=-0.8$ dex/kpc in the core region
($R<0.5$ kpc), so that the negative metallicity gradient is not
limited to the bulge region, but it continues in the core region
increasing in modulus. These values are in good agreement with the
value found from spectroscopy by M95 and by the new IR data from FTK99
who find a value of $-0.064 \pm 0.012$ dex/degree for the minor axis
of our Bulge.

\subsection{Gas and cloud phases}

We evaluate the time evolution of atomic and molecular
gas for bulge and core zones respectively. At present
time atomic and molecular gas are almost in the same quantity in the bulge
region, while in the core region molecular gas exceeds atomic gas
as from Table 3 where we
compare the total abundances for atomic and molecular gas with
observational data from Sanders, Salomon \& Scoville (1984) referring
to the zones with distances from the Galactic center $R<0.7$ kpc and
$0.7<R<1.5$ kpc. The theoretical amount of molecular gas in the core
region provides an underestimate to the observed quantity.
On the other hand the total quantity
of molecular gas estimated in both bulge and core regions together is
smaller but in reasonable agreement with observational data.
Our model takes into account infall of the atomic gas while molecular gas infall
is not considered. The great concentration of molecular gas could only be
reproduced if a similar infall parameter for the clouds in the bulge
region is introduced.

We must remind that a difference intervenes however
from the  nature of clouds: in the bulge region, clouds have
larger densities and smaller dimensions (Blitz et al. 1993) due to the
gravitational potential and/or instabilities. The central
half kpc region harbors a variety of phenomena unique in the whole galaxy.
A number of dissipative processes lead to concentration of gas into a ``Central Molecular
Zone'' (CMZ) of about 200 pc radius, containing $5 - 10 \times 10^7 M_{\odot}$ of
molecular gas. In the CMZ molecular medium is characterized by large densities
($n \ge 10^4 cm^{-3}$) and a large filling factor ($f \ge 0.1$). Such densities,
usually found only in the molecular clouds cores, are required to make clouds survive
to tidal disruption. The clouds in CMZ also present high temperatures, in the interval
30 -- 200 K, typically $\sim 70$ K, large velocity dispersion (15--50 km/s), and
apparently high magnetic fields (for a complete review see Morris \& Serabyn
1996). The bar potential and the strong tidal forces may enhance clouds velocity
dispersion, so that the dominant star formation process may be via relatively violent
events, which cause strong compression of the already dense Galactic center clouds
(Rich \& Terndrup 1997).

{psfile=fig9a.ps angle=270 hscale=40 vscale=40 hoffset=-20 voffset=+250}

\begin{table}
\caption{Gas Phases}
 \begin{tabular}{@{}lrrr@{}}  
 & & & \\
 & {H$_{2}$}   & {HI} & {H$_{2}$/HI} \\
 & ($10^7 M_{\odot}$) & ($10^7 M_{\odot}$) & \\  
 & & & \\
bulge & 2.3 & 2.50  & 0.9 \\
core  & 0.1 & 0.07 & 1.4 \\
R$<0.7$ & 3  &   &         \\
$0.7<$R$<1.5$ & $\sim$ 1 &   &  \\
R$<1.5$ & 3 - 5 & 0.10 & 50.0  \\ 
 & & & \\
\end{tabular}
\end{table}

\section{Bulges of External Galaxies}

\subsection{Theoretical Models}

Theoretical models have been calculated for bulges of nine spiral
galaxies: NGC~224, NGC~628, NGC~3198, NGC~6946, NGC~598, NGC~300,
NGC~4321, NGC~4303, and NGC~4535.
These spiral galaxies have different Hubble type, implying
that the corresponding bulges have different size and/or mass. Their
contribution to the total light of the galaxy in the B-band is higher
for early types and lower for late types.
For later galaxies we will consider
as {\sl the bulge} the central region of the disc, even if it has not
a clear bulge appearance.

The model we apply is the same used for the Galactic bulge and
described in section 2 but with a simplification: we do not consider
the presence of the core component.  Features for every galaxies used in our
models are summarized in Table 4. For every galaxy, named in column
(1), we have: the distance D in column (2), the Hubble type T in column
(3), taken from Tully (1988) and Simien \& de Vaucouleurs (1986); in
column (4), the arm class, taken from Elmegreen \& Elmegreen (1987)
and Biviano et al.  (1991); in column (5) we give the effective radius
of the disc R$_{eff}$ in kpc and in column (6) the maximum rotational
velocity.

\begin{table}
\caption[]{Observational characteristics.\label{Table 4}}
\begin{tabular}{lccccr}\\  
 & & & & & \\
Galaxy & D & T &  Arm  & R$_{eff}$  & V$_{rot,max}$ \\
 Name & (Mpc) &  & Class& (kpc) &(km/s) \\   
 & & & & & \\
NGC~224  & 0.7 & 3 & 12  &  7.8  &  250  \\
MWG      & --   & 4 & --  &  6.2  &  220  \\
NGC~4303 & 16.8 & 4 &  9  &  4.6  &  150  \\
NGC~4321 & 16.8 & 4 & 12  &  8.5  &  270  \\
NGC~628  & 7.2  & 5 &  9  &  5.0  &  220  \\
NGC~3198 & 9.6  & 5 & --  &  4.5  &  160  \\
NGC~4535 & 16.8 & 5 &  9  &  9.1  &  210  \\
NGC~598  & 0.7 & 6 &  5  &  2.6  &   85  \\
NGC~6946 & 5.7  & 6 &  9  &  5.1  &  180  \\
NGC~300  & 1.6 & 7 &  5  &  2.4  &   80  \\   
 & & & & & \\
\end{tabular}
\end{table}

The total masses are obtained
from the rotation curves referenced in MFD96 for six galaxies
(NGC~224, NGC~628, NGC~3198, NGC~6946, NGC~598, and NGC~300) and in
MHB99 for the three Virgo galaxies NGC~4321, NGC~4303 and
NGC~4535. Geometrical and mass data used as
input in our models are shown in Table 5. Discs and bulges radii are in
columns (2) and (3); the bulge and  total masses are in
columns (4) and (5).

\begin{table}
\caption[]{Input mass and geometry data.\label{Table 5}}
\begin{tabular}{lcrrr}\\   
 & & & & \\
Galaxy  & R$_{disc}$ & R$_{B}$ & M$_{B}$& M$_{tot}$ \\
 Name & (kpc) & (kpc) &  (10$^{9} M_{\odot}$) &(10$^{11} M_{\odot}$) \\  
 & & & & \\
NGC~224 & 25 & 4.0 & 40 & 4.35 \\
MWG     & 20 & 2.0 & 18 & 3.30 \\
NGC~4303& 14 & 2.0 &  8 & 1.60 \\
NGC~4321& 16 & 3.0 & 30 & 4.00 \\
NGC~628 & 16 & 2.0 & 10 & 3.00 \\
NGC~3198& 12 & 1.5 &  7 & 1.80 \\
NGC~4535& 16 & 2.0 &  9 & 3.30 \\
NGC~598 &  9 & 0.5 & 0.8& 0.50 \\
NGC~6946& 12 & 1.0 &  6 & 2.25 \\
NGC~300 &  7 & 0.1 & 0.1& 0.45 \\   
 & & & & \\
\end{tabular}
\end{table}

In Table 6, model input parameters are given:
column (2) solar equivalent radius $\rm{R_{0}}$, column (3) disc scale length
$\rm{R_{s}}$, column (4) collapse time $\tau(\rm R_{0})$; columns (5) and (6)
$\epsilon_{\mu}$ and $\epsilon_{H}$, efficiencies respectively for cloud
formation and cloud--cloud interaction, dependent on Hubble type.

\begin{table}
\caption{Model input parameters.\label{Table 6}}
\begin{tabular}{lccrcc}\\   
 & & & & & \\
Galaxy & R$_{0}$ & R$_{s}$&  $\tau_{coll}$
& $\epsilon_{\mu}$ & $\epsilon_{H}$\\
 Name & (kpc) & (kpc) & (Gyr)& &   \\   
 & & & & & \\
NGC~224 & 10 & 5.4 &  3 & 0.45& 0.50 \\
MWG     &  8 & 4.0 &  4 & 0.15& 0.08 \\
NGC~4303&  6 & 3.0 &  8 & 0.20& 0.015 \\
NGC~4321&  9 & 5.0 &  4 & 0.45& 0.10 \\
NGC~628 &  8 & 4.0 &  4 & 0.25& 0.01 \\
NGC~3198&  4 & 2.7 &  5 & 0.20& 0.02 \\
NGC~4535&  6 & 2.7 &  5 & 0.25& 0.02 \\
NGC~598 &  3 & 1.7 &  8 & 0.05& 0.005\\
NGC~6946&  5 & 4.5 &  6 & 0.18& 0.02 \\
NGC~300 &  2 & 1.7 & 15 & 0.07& 0.007\\   
 & & & & & \\
\end{tabular}
\end{table}

\subsection{Star Formation Histories}

In this section we analyze the model results for the nine studied
cases of bulges or central regions, as for late type galaxies.

We start by analyzing the star
formation related results, summarized in Table 7.
As expected more massive and early type galaxies present a more intense initial
episode of star formation, suggesting a mini--burst (see Elmegreen 1999), as shown
in figure 9.

\begin{table*}
 \centering
 \begin{minipage}{140mm}
\caption[]{Bulge model results.\label{Table 7}}
\begin{tabular}{lccccccc}\\   
 & & & & & & & \\
Galaxy  & t$_{m}$& $\Psi_{m}$ & $\Psi_{n}$
&M$_{HI}$& M$_{H_{2}}$&M$_{s}$& M$_{H}$/M$_{B}$ \\
Name& (Gyr)& (M$_{\odot}$/yr) &(M$_{\odot}$/yr)
 & (10 $^{9} \rm{M}_{\odot}$)&(10 $^{9} \rm{M}_{\odot}$)
 &(10 $^{9} \rm{M}_{\odot})$ & \\    
 & & & & & & & \\
NGC~224 &0.40& 29.2&   0.38&  0.30&  0.30& 37.50 &0.05\\
MWG     &0.60&  8.6&   0.22&  0.19&  0.20& 17.87 &0.09\\
NGC~4303&0.90&  2.4&   0.13&  0.12&  0.31&  6.88 &0.09\\
NGC~4321&0.45& 17.3&   0.30&  0.20&  0.39& 27.16 &0.08\\
NGC~628 &0.75&  4.1&   0.14&  0.09&  0.44&  8.83 &0.07\\
NGC~3198&0.85&  1.9&   0.12&  0.08&  0.19&  5.98 &0.12\\
NGC~4535&0.65&  3.7&   0.13&  0.07&  0.30&  8.00 &0.09\\
NGC~598 &0.75&  0.15&   0.02&  0.01&  0.03&  0.03 &0.50\\
NGC~6946&0.60&  1.5&   0.10&  0.03&  0.10&  4.62 &0.25\\
NGC~300 &0.40&  0.05&   0.002&  0.0002&  0.0008&  0.024&3.20\\   
 & & & & & & & \\
\end{tabular}
\end{minipage}
\end{table*}

The time when the maximum of the star formation occurs --column (2) of
Table 7-- is approximately the same for all bulges, being between 0.4
and 0.9 Gyr, always before 1 Gyr. Logically, the absolute value
maximum of the star formation rate $\Psi_{m}$, column (3), is larger
for massive bulges than for those with lower total mass. Therefore,
the stellar mass created in this early time step is larger for more
massive bulges corresponding to earlier type galaxies.
One important point is that  differences between  maximum and  present values
for the star formation rates -- columns (3) and (4)  of Table 6 -- are much lower in
panel b) than in panel a) of Figure 9. This means that the SFR is more
continuous, with a more constant value in time for bulges of later type
galaxies.  In any case, the enhancement of star formation  for the bulge
bursts are not very high in comparison with higher values reached in elliptical
or starburst galaxies.

\begin{figure*}
\vspace*{20cm}
\includegraphics{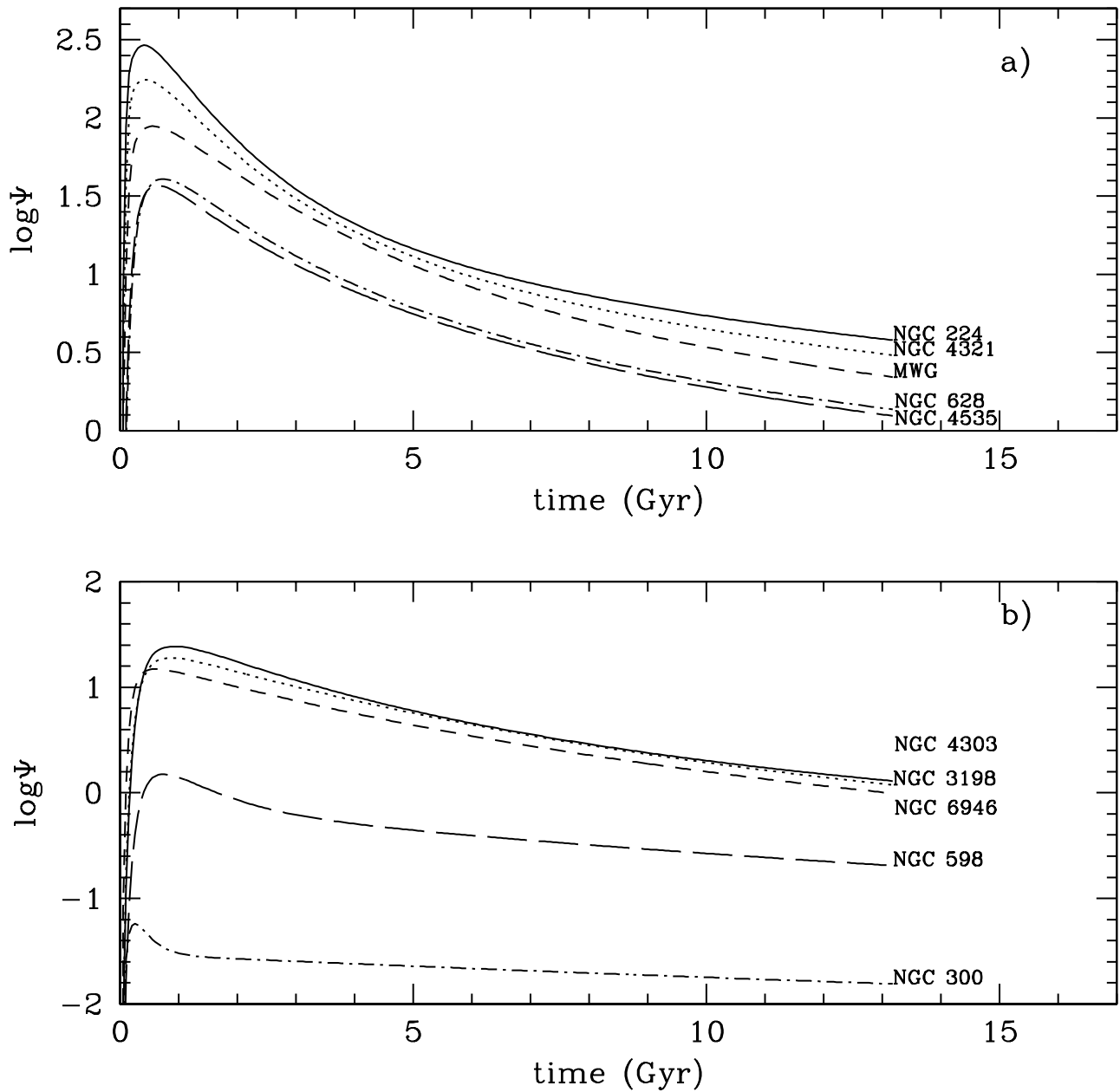}
\caption{Star formation histories predicted for our bulge sample:
a)for NGC~224, NGC~4321, MWG, NGC~618 and  NGC~4535 bulges; b) NGC~4303,
NGC~3198, NGC~6946, NGC~598 and NGC~300 bulges.\label{SFH}}
\end{figure*}


The ratio between past and present values is correlated with the Hubble
type or with the Arm Class, since the ratio $\Psi(m)/\Psi (n)$
varies with continuity.
Indeed the morphological classification of galaxies in Hubble types is related to the
bulge -- disc luminosities ratio; a correlation
between Hubble type and SFR in the bulge is expected due to
the existent correlation between total mass and Hubble type. We quantify this
relationship, that must be reproduced by coherent models of
galaxy evolution.

From our model, early type galaxies have larger and more massive bulges and
form a large number of stars in the first phase of the starburst that interests
the bulge region. There also exist a correlation
between the star formation rate in the bulge and the Arm Class of the disc;
this means that the star formation of the bulge depends also on
the dynamical features of the whole galaxy, as we underlined in previous
papers where we analyzed the correlation of galactic model properties,
their observational counterparts and the corresponding Arm Class classification
scheme, viewed as directly related to large scale dynamical processes.


\subsection{Gas Masses}

The star formation processes consume the gas in the bulge; both phases
are larger in absolute value of mass -- columns (5) and (6) of
Table 6 -- in earlier type bulges, but this effect is due to the higher
values of total mass. Taking into account geometrical effects, we
calculate surface densities for atomic and molecular gas; it results
that bulges of later type maintain higher values for both
 atomic and molecular gas surface densities:
earlier bulges consumed their gas more rapidly than the later
ones.
The total molecular gas quantity depends on two process: the
consumption to form stars and the formation of molecular
clouds from diffuse gas. Indeed, the ratio H$_{2}$/HI has a peculiar behavior:
it is lower for early type bulges because the molecular gas has been converted
in stars, but it is also low for later bulges due to the slow process of
molecular cloud formation. Only intermediate type bulges, such those with T=5
if massive enough, have high values for the ratio H$_{2}$/HI, which
may results five times higher than in T=3 type.
It is interesting to note that this trend
is completely different for spiral discs, where observed H$_{2}$/HI quantity
is larger for early type galaxies (Deveraux \& Young, 1989), as also derived in
MFD96. The same kind of behaviour occurs when the ratio H$_{2}$/HI is
represented versus the Arm Class: only intermediate galaxies
show higher proportions of molecular clouds in their bulges.


The total stellar mass of bulge stars -- given in column (7) of
Table 6 -- is decreasing with the Hubble type. For less massive
bulges the collapse lasts a longer time, and a larger proportion of mass
remains in the halo zone such it can be seen in column (8). Moreover the
already collapsed mass needs a longer time to form stars because the
efficiencies to form clouds and stars are lower.

\subsection{Gas Chemical Abundances}

\begin{table*}
 \centering
 \begin{minipage}{140mm}
\caption{Chemical Evolution  Results.\label{Table 8}}
\begin{tabular}{lccccccc} \\   
 & & & & & & & \\
Galaxy  & [O/H]$_{m}$& [Fe/H]$_{m}$& [Mg/Fe]$_{m}$&
 [O/H]$_{n}$& [Fe/H]$_{n}$& [Mg/Fe]$_{n}$ &Z$_{n}$\\    
 & & & & & & & \\
 NGC~224   &  -0.61 &  -0.77 &   0.33 &  -0.01  &   0.37 &  -0.12 &  0.029 \\
 MWG       &  -0.54 &  -0.70 &   0.33 &   0.00  &   0.31 &  -0.05 &  0.030 \\
 NGC~4303  &  -0.71 &  -0.77 &   0.23 &  -0.02  &   0.27 &  -0.05 &  0.025 \\
 NGC~4321  &  -0.59 &  -0.76 &   0.33 &   0.00  &   0.33 &  -0.08 &  0.028 \\
 NGC~628   &  -0.68 &  -0.74 &   0.23 &   0.00  &   0.30 &  -0.06 &  0.023 \\
 NGC~3198  &  -0.61 &  -0.67 &   0.23 &  -0.04  &   0.25 &  -0.05 &  0.024 \\
 NGC~4535  &  -0.67 &  -0.71 &   0.34 &  -0.02  &   0.29 &  -0.06 &  0.026 \\
 NGC~598   &  -0.78 &  -0.77 &   0.25 &  -0.14  &   0.11 &  -0.03 &  0.019 \\
 NGC~6946  &  -0.47 &  -0.62 &   0.28 &  -0.03  &   0.26 &  -0.07 &  0.025 \\
 NGC~300   &  -0.63 &  -0.86 &   0.40 &  -0.12  &   0.11 &  -0.01 &  0.019 \\ 
 & & & & & & & \\
\end{tabular}
\end{minipage}
\end{table*}

We summarize our results concerning heavy element abundances in Table 8.
A few aspects deserve a detailed discussion; in columns (3), (4), and (5)
we show oxygen and iron abundances relative to hydrogen and magnesium over
iron abundances at the time of SFR maximum.
Since the bulk of stars created near that time, the
average abundances of the corresponding bulge populations are reasonably
well described by these values. Due to the
differences in the star formation histories, we could imagine a priori
that every bulge has a different metallicity.
However, the total abundances at the SF maximum turn out
very similar for all bulges, with values $\sim \Zsun/10$. The oxygen
abundances also are very similar, around $-0.60$ dex, while the iron abundance
is lower, approximately around $-0.75$ dex (it is synthesized mostly by SN
Type I, exploding with a delay with respect to the main episode of SF, and
a dispersion  exists depending on the total mass in the bulge).
Therefore the oldest stellar generation  with the
largest number of stars have approximately the same abundance for every bulge,
independently of the Hubble type.

For the present time, we have the results of columns (5) [O/H]$_{n}$, (6)
[Fe/H]$_{n}$, (7) [Mg/Fe]$_{n}$ and (8)$Z{_n}$. The total abundance almost
reaches \zsun\ in all cases with a weak correlation between $Z_{n}$ and the
mass of the bulge, the Hubble type or the Arm Class. However there are
differences between elements. The oxygen abundance is about solar in all cases
showing a saturation level for all bulges, but the iron abundance is higher
for earlier Hubble types.


The average abundances for
stellar populations, calculated taking into account
all generations of stars are reproduced  in Table 9 where
column (2) is the average total abundance $\overline {\rm Z}$, (3) average
oxygen abundance $\overline {[\rm O/H]}$, (4) average calcium abundance $\overline {[\rm Ca/H]}$,
(5) average iron abundance $\overline {[\rm Fe/H]}$ and (6) average magnesium to iron
ratio $\overline {[\rm Mg/Fe]}$.
The stellar abundances are as expected intermediate between maximum and
present day gas abundances. The only possible comparison with data
concerning MWG is certainly positive, since the mean stellar
metallicity is $\overline {\rm [Fe/H]}=-0.17$;
this value is in agreement with the most recently
estimates from FTK99, who give a value of $-0.2$ dex and also with
McWilliam \& Rich (1994)($-0.25$ dex) and the other author's data
referenced in the Introduction.  Metallicities predicted for our
bulges show subsolar values for the stellar populations and hence seem to
exclude any similarity to massive metal-rich elliptical galaxies.

The different nucleosynthetic origin of  $\alpha$-elements and  iron
is evident in the behavior of  [Mg/Fe],  higher than solar for older stellar
populations. In Table 7 the value reached at $t_{m}$ is $[\rm Mg/Fe] \sim 0.2$
dex, with a correlation with the Hubble type;
[Mg/Fe] takes the highest values at ${t_m}$, while at the
present time it is below solar. At ${t_m}$ larger values are for
earlier bulges, while at present larger values are for later bulges.
The conspiracy of SF and nucleosynthesis from different populations
determine the average stellar abundance to follow a smooth variation
with type.

\begin{table}
\caption{Average Stellar Abundances.}
\begin{tabular}{lccccc}\\    
 & & & & & \\
Galaxy&  $\overline {\rm Z} $&  $\overline {\rm [O/H]}$ & $\overline {\rm[Ca/H]}$ &
$\overline {\rm [Fe/H]} $ & $\overline {\rm[Mg/Fe]}$ \\   
 & & & & & \\
NGC~224 &  0.015 & -0.31 & -0.19 & -0.22 & 0.11 \\
MWG     &  0.015 & -0.29 & -0.17 & -0.17 & 0.08 \\
NGC~4303&  0.014 & -0.36 & -0.24 & -0.24 & 0.07 \\
NGC~4321&  0.015 & -0.30 & -0.18 & -0.21 & 0.10 \\
NGC~628 &  0.013 & -0.37 & -0.26 & -0.27 & 0.09 \\
NGC~3198&  0.014 & -0.31 & -0.19 & -0.17 & 0.06 \\
NGC~4535&  0.014 & -0.34 & -0.22 & -0.23 & 0.08 \\
NGC~598 &  0.013 & -0.34 & -0.21 & -0.20 & 0.06 \\
NGC~6946&  0.016 & -0.24 & -0.11 & -0.09 & 0.05 \\
NGC~300 &  0.016 & -0.23 & -0.10 & -0.07 & 0.04 \\   
 & & & & & \\
\end{tabular}
\end{table}

\section{Stellar Populations and Spectral Indices}

Once evaluated the global evolution of the bulge region,
 in particular the history of star formation, and hence
the sequence of stellar generations and their corresponding
element abundances, we may determine the spectral indices for
the stellar populations, adopting the synthesis model results
for the Single Stellar Populations (SSP). This method goes in the
reverse direction of other authors (see e.g. Jablonka et
al. 1996).

We use the fitting functions given by Worthey et al. (1995)
to assign the spectral indices $\rm Mg_{2}$, $\rm H\beta$ and
$\rm <Fe>$ to each SSP, building blocks of the bulges, as functions of
age and metallicity.
We also assign a continuum flux, calculated with a new synthesis model
developed by Moll\'a et al. (1999), and finally
we combine all generations in the proportion determined by the
created stellar mass at every time step (or at every age).

Results for the studied bulges are given in Table 10; in Figure
10 we show our predictions in the plane $\rm Mg_{2}\ - <Fe>$ with a set
of spectral indices data observed in bulges by Jablonka et al. (1996), Idiart
et al (1996a,b), Beauchamp \& Hardy(1997), 
Vazdekis et al. (1997) and Samson et al. (1998),
and the data from Gorgas et al. (1997) for low mass
spheroidal galaxies.  Our theoretical predictions are
in agreement with the data observed in similar regions and with
theoretical model for SSP obtained by Worthey (1994) and by Borges et
al. (1995) and Idiart \& Freitas--Pacheco (1995).
Spheroidal galaxies have spectral indices very similar to those of
bulges. Galaxies with large values for both indices are out of the
theoretical lines, probably in agreement with larger
[Mg/Fe].

First we determined the star formation and chemical evolution,
then the spectral properties, as a direct consequence of both.
The advantages in affording the problem with this strategy are
evident in  comparison with
the standard analysis, based on the fitting procedure.
In our study, from the first principles, late type
galaxies form most stars when  [Mg/Fe] has decreased in
comparison with large bulges where stars are principally formed at
early times, with high [Mg/Fe] values. Idiart et al. (1996b),
to reproduce observations, require
[Mg/Fe] values sensibly higher than ours, with [Mg/Fe] reaching 0.6 dex.
Our models have a mean stellar [Mg/Fe] between 0.11 and 0.05.
The multiphase model gave a mean stellar bulge metallicity $\overline {\rm
[Fe/H]} \sim -0.2$, in agreement
with all recent estimations. If the
statement of  Peletier \& Bacells  (1996) that all bulges are similar in
metallicities is true, our model is in agreement with them: our
results of Table 9 give $\overline {\rm Z} \sim 0.015$ and
$\overline {\rm [Fe/H]} \sim -0.20$.
Idiart et al. (1996a) for the Galactic bulge require
$\overline {\rm [Fe/H]} = -0.19$, but Idiart et
al. (1996b) obtain $\overline {\rm [Fe/H]}=-0.02$ (almost solar) for the
bulges of external galaxies. This difference is striking and we cannot
explain why those models result so different for the
Galactic bulge and the others. In our case all bulges are similar in
abundances.

\begin{figure*}
\vspace*{10cm}
\includegraphics{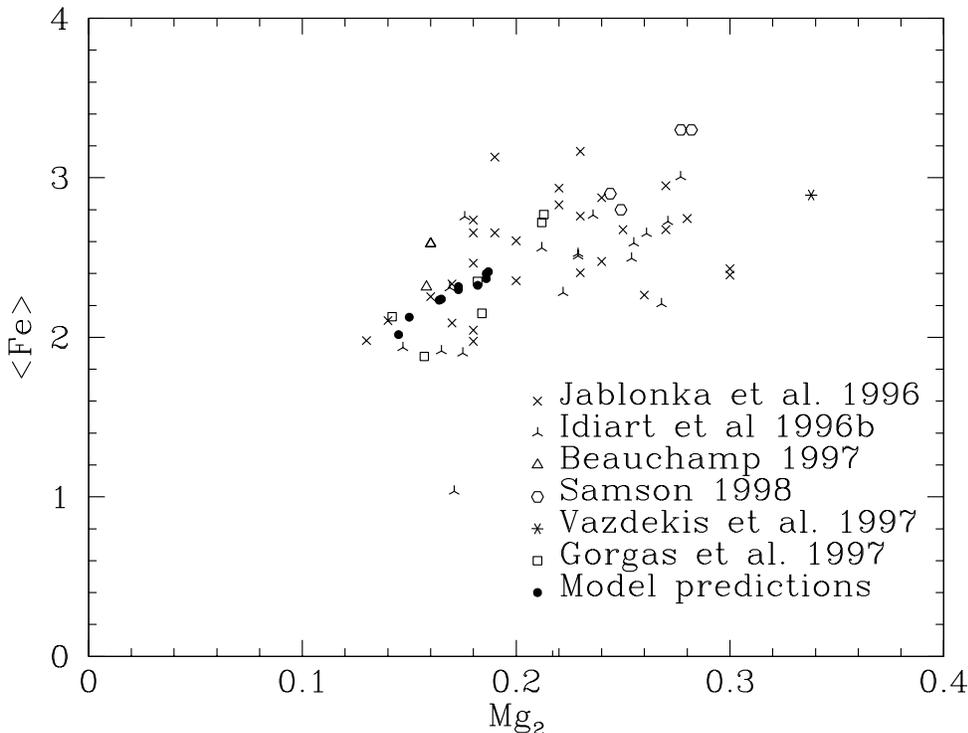}
\caption{Spectral index Mg$_2$ versus Fe52.}
\end{figure*}

\begin{table}
\caption[]{Predicted Spectral Indices.\label{Table 10}}
\begin{tabular}{lccc}\\    
 & & &  \\
Galaxy&  $Mg_{2}$ & Fe52 & H$\beta$\\
name  & (mag) &  (\AA) & (\AA) \\
 & & &  \\
NGC~224 &  0.19 & 2.51 & 2.32 \\
MWG     &  0.19 & 2.54 & 2.41 \\
NGC~4303&  0.16 & 2.36 & 2.45 \\
NGC~4321&  0.18 & 2.48 & 2.05 \\
NGC~628 &  0.17 & 2.37 & 2.38\\
NGC~3198&  0.17 & 2.43 & 2.70\\
NGC~4535&  0.17 & 2.45 & 2.33\\
NGC~598 &  0.15 & 2.25 & 3.11\\
NGC~6946&  0.19 & 2.54 & 2.14\\
NGC~300 &  0.14 & 2.13 & 3.49\\   
 & & &  \\
\end{tabular}
\end{table}

The spectral indices Mg$_{2}$ and $<\rm Fe>$ show different dependence on
the morphological type and/or the total mass of the bulge:
Mg$_{2}$ depends on T while $<\rm Fe>$ maintains constant. This results
was already shown by Jablonka et al. (1996), when they represented
their observed indices as a function of the luminosity M$_{r}$. The
luminosity has a correlation with the mass of the bulge, which in turn
depends on the total mass of the galaxy and on the morphological type.
Therefore, a clear dependence between the spectral index Mg$_{2}$ and
T is expected. The index $<\rm Fe>$ does not shown any trend with T,
neither with M$_{r}$. This different behavior is not explained by the
different mean value [Mg/Fe], neither by differences between iron
abundances in the stellar populations of different morphological
types, but by the different mean ages of their populations. The
earlier galaxies have old stellar populations with large magnesium
abundances. Both effects increase the values of Mg$_{2}$. Later
type galaxies have young populations in larger proportions. These
populations have low indices and the higher proportion in mass and
luminosity in the integrated quantities decrease the values of indices
Mg$_{2}$. For the index $<\rm Fe>$ the effect of the ages is the same, but
iron abundances increase for later types, which produce more iron when
stars are creating. Therefore, the integrated $<\rm Fe>$ is due to a
mixing of old and metal--poor stellar populations and these young
populations rich in iron in the later type galaxies, while in the
earlier proceed only from old stellar populations. Both effect may
produce similar results, and in consequence the final $<\rm Fe>$ does not
depend on the morphological type or the total luminosity.



\section{Discussion and Conclusions}

The multiphase model, inserted in the dissipative collapse scenario
following ELS picture, is able to reproduce the observed
properties of bulges. The determination of the history of star formation
as the principal ingredient of the evolution is useful in the
understanding of the bulge variations in the Hubble sequence.
In massive bulges corresponding to early type galaxies, the intensity
of the star formation maximum is
large enough to consume the whole gas, by preventing the subsequent
star formation. The young stellar populations do not exist or they
exist in very small proportions in comparison with the bulk of stars
formed in earlier times. In later type galaxies, less massive, this
maximum is not so large, and the star formation rate has
almost a constant intensity. These results are in
agreement with observations in the K-band from Seigar \& James
(1998a,b), whose results seem to indicate that the morphological types
must be controlled by the star formation ratio between past and
present and not by the ratio bulge/disc, because it does not present
any correlation with the morphological type T.

Chemical results show bulges very similar in their abundances at the
present time and for the time when the star formation maximum occurs,
in agreement with estimations obtained from colors data (Peletier \&
Balcells, 1996; deJong, 1996a,b). Mean abundances have also been
calculated by showing subsolar values  and
similar for all bulges independently of the Hubble type, the arm
class and/or the luminosity of their host galaxies.

With a method reverse to the one adopted by Jablonka et al.
(1996) and other authors, we compute the spectral indices Mg$_{2}$ and
$<\rm Fe>$ from our stellar generation results, by using SSP synthesis
models. Our results are in agreement with a set of
observations for bulges in the plane Mg$_{2}-<\rm Fe>$.
We may explain why the first of this indices depend on the
Hubble type or the total luminosity of the bulge, while the index $<\rm
Fe>$ does not show a similar trend. 
The evolution and the star formation histories of our
bulge models are crucial for this respect: 
 the mean age of the stellar populations
in bulges are different, even if the formation of the bulge himself
starts at the same moment. The star formation is more continuous in
the central regions of the late type galaxies than in the early
type ones, by producing more iron in the stars of the first ones.
This mixing of the old metal poor populations and young and iron rich
populations in the late type galaxies produce the same result for
$<\rm Fe>$ than is produced by the old metal poor population created
in the early type galaxies in the first phases of star
formation. The magnesium is produced instead very quickly in these
last galaxies, by increasing the abundance of the stars. Therefore the
index Mg$_{2}$ show a clear trend with the morphological type because
both ages and abundances have the same effect, by increasing the value
of  Mg$_{2}$.

The model shows that it is not needed the consider bars or radial flows to
explain the features of bulges or their formation.
Theoretical results for intermediate type bulges show the highest ratio
H$_{2}$/HI. These galaxies present
the strongest observed bars, as from K band data; this suggests
that bars may appear and disappear in the
evolution of the bulge. They would contribute to produce a new starburst phase
in bulges, but it may occurs several Gyr after the formation of the
bulge. In this case, the gas flows into the central region, and
younger stars will be created in the bulge. Their effects would be
measured if the quantity of inflow gas is large enough or comparable
to the mass of stars already created in the bulge. Otherwise, their
influence will be negligible in the indices
Mg$_{2}$ and $<\rm Fe>$ and we could only reveal them if we use other
kind of data such the calcium triplet or H$\beta$.

We conclude presenting our answer to the questions addressed in the
introduction:
\begin{itemize}
\item{} it is not needed to require the merging and/or accretion of
external material to reproduce the main observed characteristics of
bulges; the self--collapse of the protogalaxy seems to be sufficient
to give the correct time scales, enrichment history and populations
to reproduce observations not only for the Milky Way but also for
external spiral galaxies.

\item{} The observed properties of bulge
stellar populations are well reproduced in detail if a correct evaluation
of the contribution from the various regions involved in the bulge
formation and evolution are considered, we mean the halo, bulge and core
populations.

\item{} There is no evidence of an analogy with elliptical galaxies: the
apparent geometrical similarity does not correspond to any functional
and physiological equivalence between these two families of astronomical
objects.

\item{} The mean age of bulges varies according to the galactic type of the
parent galaxy: there is a strong correlation between the bulge populations
and the whole galactic evolution.

\end{itemize}

In conclusion, not only it makes sense to figure bulge and disc evolution as
two joint components of an intrinsically unitary structure,
along the Hubble sequence, but it appears to be the only realistic way
to obtain a good comprehension of bulge structure.

\section*{Acknowledgments}
M. Moll\'{a} acknowledges the Spanish {\sl Ministerio de Educaci\'{o}n y
Cultura} for its support through a post-doctoral fellowship.
F. Ferrini acknoledges the MURST for financial support.
This research has made use of the
NASA/IPAC Extra-galactic Database (NED), which is operated by the Jet
Propulsion Laboratory, Caltech, under contract with the National
Aeronautics and Space Administration; the NASA's Astrophysics Data System
Abstract Service; and the automated archive astro-ph, which is supported
by the U. S. National Science Foundation under agreement with Los Angeles
National Laboratory and by the U.S Department of Energy.

\bsp
\label{lastpage}
\end{document}